\documentclass[11pt]{article}
\usepackage[a4paper, margin=1in]{geometry}
\usepackage{graphicx}
\usepackage{color}
\usepackage{todonotes}

\newcommand{\ytodo}[1]{\todo[color=blue!20, inline]{Yasamin:#1}}

\definecolor{darkgreen}{rgb}{0,0.5,0}
\usepackage{hyperref}
\hypersetup{
    unicode=false,          
    colorlinks=true,        
    linkcolor=red,          
    citecolor=darkgreen,    
    filecolor=magenta,      
    urlcolor=cyan           
}

\usepackage{amsthm, amsmath, amssymb, amsfonts, mathrsfs, mathtools, cite}
\usepackage{verbatim}
\usepackage{footnote}
\usepackage[linesnumbered,boxed, vlined]{algorithm2e}
	\DontPrintSemicolon
  \SetKwInOut{Input}{Input}
  \SetKwInOut{Output}{Output}
\usepackage{lineno}
\usepackage{caption}
\usepackage{framed}
\usepackage[framemethod=tikz]{mdframed}
\global\mdfdefinestyle{myframe}{leftmargin=.75in,rightmargin=.75in,linecolor=black,linewidth=1.5pt,innertopmargin=10pt,innerbottommargin=10pt} 
\usepackage{enumerate}

\usepackage{tikzsymbols}
\usepackage{thmtools,thm-restate}
\usepackage{nicefrac}

\usepackage[capitalize, nameinlink]{cleveref}
\Crefname{theorem}{Theorem}{Theorems}
\Crefname{lemma}{Lemma}{Lemmas}
\Crefname{claim}{Claim}{Claims}
\Crefname{observation}{Observation}{Observations}
\Crefname{algorithm}{Algorithm}{Algorithms}
\Crefname{enumi}{Step}{Steps}

\DeclareMathOperator{\poly}{poly}

\theoremstyle{plain}
\newtheorem{theorem}{Theorem}[section]
\newtheorem{lemma}[theorem]{Lemma}
\newtheorem{corollary}[theorem]{Corollary}
\newtheorem{definition}[theorem]{Definition}

\newcommand{\eps}{\varepsilon}

\newcommand{\rb}[1]{\left( #1 \right)}
\newcommand{\ee}[1]{\mathbb{E} \left[ #1 \right]}
\newcommand{\prob}[1]{\Pr \left[ #1 \right]}

\newcommand{\tO}{\tilde{O}}

\newcommand{\cC}{\mathcal{C}}
\newcommand{\cc}{\mathbf{c}}
\newcommand{\pc}{p_{S}}
\newcommand{\cF}{\mathcal{F}}
\newcommand{\cD}{\mathcal{D}}
\newcommand{\cE}{\mathcal{E}}
\newcommand{\rE}{\mathsf{E}}
\newcommand{\cN}{\mathcal{N}}
\newcommand{\cP}{\mathcal{P}}
\newcommand{\cR}{\mathcal{R}}

\newcommand{\clique}{\textsc{Congested Clique}\xspace}

\newcommand{\eqdef}{\stackrel{\text{\tiny\rm def}}{=}}

\newcommand{\patharrow}{\raisebox{0.5ex}{\tikz \draw[->, line width=0.2ex] (0,0) -- (0.5,0);}}

\newcommand{\MPC}[0]{\ensuremath{\mathsf{MPC}}}

\newcommand{\PRAM}[0]{\ensuremath{\mathsf{PRAM}}}

\newcommand{\remove}[1]{}

\title{Massively Parallel Algorithms for Distance Approximation and Spanners}
\author{Amartya Shankha Biswas \\ CSAIL, MIT
\and Michal Dory \\ ETH Zurich
\and Mohsen Ghaffari\\ ETH Zurich
\and Slobodan Mitrovi\' c \\CSAIL, MIT
\and Yasamin Nazari\\ Johns Hopkins University}

\begin{document}
\date{}
\maketitle

\begin{abstract}
Over the past decade, there has been increasing interest in distributed/parallel algorithms for processing large-scale graphs.
By now, we have quite fast algorithms---usually sublogarithmic-time and often $\poly(\log\log n)$-time,
or even faster---for a number of fundamental graph problems in the massively parallel computation ($\MPC$) model.
This model is a widely-adopted theoretical abstraction of MapReduce style settings,
where a number of machines communicate in an all-to-all manner to process large-scale data.
Contributing to this line of work on $\MPC$ graph algorithms,
we present $poly(\log k) \in \poly(\log\log n)$ round $\MPC$ algorithms for computing $O(k^{1+{o(1)}})$-spanners
in the strongly sublinear regime of local memory. To the best of our knowledge, these are the first sublogarithmic-time $\MPC$ algorithms for spanner construction. 

\medskip
\noindent As primary applications of our spanners, we get  two important implications, as follows:
\begin{itemize}
    \item For the MPC setting, we get an $O(\log^2\log n)$-round algorithm for $O(\log^{1+o(1)} n)$ approximation of all pairs shortest paths (APSP) in the near-linear regime of local memory. To the best of our knowledge, this is the first sublogarithmic-time $\MPC$ algorithm for distance approximations.
    
    \item Our result above also extends to the \clique model of distributed computing, with the same round complexity and approximation guarantee. This gives the first \emph{sub-logarithmic} algorithm for approximating APSP in \emph{weighted} graphs in the \clique model.
\end{itemize}
\end{abstract}

\section{Introduction and Related Work}
\label{sec:introduction}
\subsection{Massively Parallel Computation}
Processing large-scale data is one of the 
indubitable necessities of the future, and one for which we will rely more and more on distributed/parallel computation. Over the past two decades, we have witnessed the emergence and wide-spread usage of a number of practical distributed data processing frameworks, including MapReduce~\cite{dean2008mapreduce}, Hadoop~\cite{white2012hadoop}, Spark~\cite{ZahariaCFSS10} and Dryad~\cite{isard2007dryad}. More recently, there has also been increasing interest in building a corresponding algorithmic toolbox for such settings. By now, there is a de-facto standard theoretical abstraction of these frameworks, known as the \emph{Massively Parallel Computation} ($\MPC$) model. The model was introduced first by Karloff et al.~\cite{karloff2010MapReduce} and refined later by Beam et al.~\cite{Beame13} and Goodrich et al.~\cite{goodrich2011sorting}. 

\subparagraph*{MPC model.} On a very high-level, the model assumes a number of machines, each with a memory capacity polynomially smaller than the entire data, which can communicate in an all-to-all fashion, in synchronous message passing rounds, subject to their memory constraints. More concretely, in the $\MPC$ model~\cite{karloff2010MapReduce,goodrich2011sorting, Beame13}, we are given an input of size $N$ which is arbitrarily distributed among a number of machines. Each machine has a memory of size $S = N^{\alpha}$ for some $0 < \alpha < 1$, known as the local memory or memory per machine. Since the data should fit in these machines, the number of machines is at least $N/S$, and often assumed to be at most $O(N/S \poly(\log N))$. Hence, the global memory---that is, the summation of the local memories across the machines--- is $\tilde{O}(N)$. The machines can communicate in synchronous message-passing rounds, subject to the constraint that the total amount of messages a machine can communicate per round is limited by its memory $S$. 
 
In the case of graph problems, given a graph $G=(V,E)$ the total memory $N$ is $O(|E|)$ words. Ideally, we would like to be able to work with machines that have a small local memory, and still have only few rounds. However, the task gets more complex as one reduces the local memory. Depending on how the local memory compares with the number of vertices $n=|V|$, there are three regimes that have been studied in the $\MPC$ model: 
\begin{itemize}
    \item the \emph{strongly superlinear} regime where $S\geq n^{1+\eps}$ for a constant $\eps>0$, 
    \item the \emph{near-linear} regime where $S=\tilde{O}(n)$, and 
    \item the \textit{strongly sublinear} regime where $S=n^{\gamma}$ for a positive constant $\gamma<1$.
\end{itemize}
We note that the algorithms that can work in the strongly sublinear memory regime are sometimes referred to as \emph{scalable massively parallel algorithms}. Our focus will be on the more stringent, and also more desirable, regimes of near-linear and strongly sublinear memory.

\subsection{Graph Problems and Massively Parallel Computation}
At the center of the effort for building algorithmic tools and techniques for large-scale data processing has been the subarea of \emph{$\MPC$ algorithms for graph problems}, e.g.,~\cite{goodrich2011sorting, LattanziMSV11, Beame13, Andoni:2014, Beame14, hegeman2015lessons, AhnGuha15,Roughgarden16,Im17,czumaj2017round, assadi2017simple,AssadiBBMS19,ghaffari2018improved,harvey2018greedy,brandt2018breaking,boroujeni2018approximating,Andoni2018, Behnezhad0DFHKU19,BehnezhadHH19,BehnezhadDELM19,Andoni2018,AndoniSZ19,AssadiBBMS19,AssadiSW19,GamlathKMS19,GhaffariLM19,GhaffariU19, LackiMOS20,ItalianoLMP19,ChangFGUZ19,GhaffariKU19,GhaffariNT20-cut, GhaffariN20-cut}. We discuss a very brief overview here. Please see \Cref{app:related} for a more detailed overview, with quantitative bounds. 

Early on~\cite{karloff2010MapReduce, Beame13}, it was observed that $\MPC$ can simulate classic $\PRAM$ parallel algorithms (subject to conditions on the total amount of work) in the same time. This immediately led to $\poly(\log n)$ round algorithms for a wide range of graph problems. 

Since then, the primary objective in the study of $\MPC$ algorithms has been to obtain significantly faster algorithms, e.g., with constant or $\poly(\log\log n)$ round complexity. This was achieved initially for the strongly super-linear memory regime, for many problems, and over the past few years, there has also been significant progress on near-linear and strongly sublinear memory regimes. In particular, there has been quite some progress for (A) global graph problems such as connected components, maximal forest, minimum-weight spanning tree, minimum cut, etc, ~\cite{LattanziMSV11, AhnGM12, Andoni2018,BehnezhadDELM19, GhaffariNT20-cut, GhaffariN20-cut} and (B) local graph problems such as maximum matching approximation, graph coloring, maximal independent set, vertex cover approximation, etc~\cite{LattanziMSV11, harvey2018greedy, czumaj2017round,ghaffari2018improved,AssadiBBMS19,AssadiCK18, ChangFGUZ19,ChangFGUZ19,RozhonGhaffari20}.

\subparagraph*{Distance Problems.}
Despite the substantial progress on various graph problems, one fundamental category of graph problems for which the progress in $\MPC$ has been slower is distance computations and, more generally, distance-related graph problems. In particular, considering that a key criteria in the area is to obtain near constant (and especially $o(\log n)$) time algorithms, the following question has remained open.

\medskip
\begin{center}
\begin{minipage}{0.95\linewidth}
    \textbf{Question:} \textit{Are there $\poly(\log\log n)$-time MPC algorithms in the near-linear memory regime that compute all pairs shortest paths, or any reasonable approximation of them? }
\end{minipage}
\end{center}

\medskip
In fact, to the best of our knowledge, prior to our work, there was no known $\MPC$ algorithm even with a sublogarithmic round complexity, for any non-trivial approximation factor, and even for single-source shortest paths. 

The known algorithms provide only $poly(\log n)$ round complexity, which is considerably above our target. For instance, one can obtain a $poly(\log n)$ round algorithm for $1+\eps$ approximation of single source shortest paths (SSSP) in the sublinear memory regime of $\MPC$, by adapting the $\PRAM$ algorithm of Cohen~\cite{cohen2000polylog}. While that algorithm requires $m^{1+\Omega(1)}$ global memory, recent $\PRAM$ results reduce that to $\tilde{O}(m)$~\cite{li2020_PRAM, AndoniCZ20}. Among more recent work, Hajiaghayi et al.~\cite{hajiaghayi2019mapreduce} give an all pairs shortest path (APSP) algorithm that runs $O(\log^2 n)$ rounds in strongly sublinear local memory, for certain range of edge weights, and uses a large polynomial global memory, which depends on exponents of matrix multiplication.

To tackle the above question, we can naturally ask whether, in the allowed time, one can sparsify the graph considerably, without stretching the distances -- as that sparse graph then can be potentially moved to one machine. This directly brings us to the notion of spanners, as introduced by Peleg and Sch{\"a}ffer \cite{peleg1989graph}, which are subgraphs with few edges that preserve distances, up to a certain multiplicative factor. A more detailed explanation follows.

\subparagraph{Spanners.}
Given a graph $G=(V,E)$, a $k$-spanner is a (sparse) spanning subgraph $H$ of $G$ such that the distance between each pair of nodes in $V$ on this subgraph $H$ is at most $k$ times their distance in the original graph $G$ \cite{Peleg00}. It is known that every graph admits a $(2k-1)$-spanner of size $O(n^{1+1/k})$, for $k \geq 1$, and assuming Erd{\"o}s girth conjecture, this tradeoff is also tight.
%
%
Spanners have found many applications in various models, such as, in constructing sparsifiers in the streaming model~\cite{kapralov2014}, designing work-efficient $\PRAM$ algorithms \cite{friedrichs2018}, and transshipment based distance approximations in $\PRAM$~\cite{li2020_PRAM} and Congested Clique~\cite{becker_et_al:transshipment}. In the context of strongly sublinear memory $\MPC$, \cite{dinitz2019} used spanners to obtain an exponential speed up in preprocessing of distance sketches (though, still requiring polylogarithmic rounds). In particular, they achieve this by using spanners in order to simulate non-work-efficient $\PRAM$ algorithms in $\MPC$ without using extra memory. In general, spanners can be applied to reduce use of resources such as communication and memory for distance-related computation on denser graphs at the expense of accuracy.

We are not aware of any $\poly(\log\log n)$ round MPC algorithm for computing spanners, in the sublinear or near-linear memory regime. Though one can obtain some (weaker) results from those of other computational models, as we will be discussed later, when mentioning our results.

\subsection{Our Contribution}
In this paper, we provide the first sublogarithmic time $\MPC$ algorithms for distance approximations and for constructing spanners. Our spanner construction works in the strongly sublinear memory regime, while for using them in distance approximation we need to move to the near-linear memory regime. Since, as stated, spanners are versatile tools that have found applications in various distance-related graph problems, we are hopeful that our spanner construction should also help in a wider range of problems. 

\paragraph{Spanner Constructions.}
Our main technical result, as stated below, provides a family of algorithms which provide a general trade-off
between the number of $\MPC$ rounds and the stretch of the resulting spanner.
\begin{restatable}{theorem}{theoremmain}
\label{theorem:main}
Given a weighted graph $G$ on $n$ vertices and $m$ edges and a parameter $t$, there is an algorithm that runs in $O(\frac{1}{\gamma} \cdot\frac{t\log k}{\log(t+1)})$ rounds of $\MPC$ and w.h.p.~outputs a spanner of size $O(n^{1+1/k} \cdot (t + \log k))$, and stretch $O(k^s)$ when memory per machine is $O(n^\gamma)$ for any constant $\gamma > 0$, and where $s= \frac{\log (2t+1)}{\log(t+1)}$. This algorithm uses total memory of $\tO(m)$.
\end{restatable}

Since the theorem statement in \Cref{theorem:main} might be complex due to the number of variables involved, we next state several interesting corollaries of this theorem.
\begin{corollary}
\label{corollary:k-log 3}
Given a total memory of $\tO(m)$ and memory per machine being $O(n^\gamma)$, for any constant $\gamma > 0$, there is an $\MPC$ algorithm that w.h.p.~outputs a spanner with the following guarantees in terms of round complexity, stretch and size:
\begin{enumerate}
    \item runs in $O(\log k)$ rounds, has $O(k^{\log 3})$ stretch and $O(n^{1+1/k} \cdot \log k)$ size;
    \label{item:one_BS_step}
    \item runs in $O\rb{2^{1/\eps} \cdot \eps \cdot \log k}$ rounds, has $O(k^{1+\eps})$ stretch and $O(n^{1+1/k} \cdot (2^{1/\eps} + \log k))$ size;
    \item runs in $O(\frac{\log^2 k}{\log \log k})$ rounds, has $O(k^{1+o(1)})$ stretch and $O(n^{1+1/k} \cdot \log k)$ size;
    \item runs in $O(\frac{\log^2 \log{n}}{\log \log \log{n}})$ rounds, has $O(\log^{1+o(1)} n)$ stretch and $O(n \cdot \log \log{n})$ size.
\end{enumerate}
\end{corollary}
 Moreover, and crucially for our distance approximation applications, our algorithms are applicable to weighted graphs.

$\PRAM$ algorithms for spanners such as Baswana-Sen \cite{baswana_sen} have depth at least $\Omega(k)$, which leads to $\MPC$ algorithms with $O(k)$ complexity. Our algorithms are significantly faster, at the price of a small penalty in the stretch.
Another relevant prior work that we are aware of is a one of Parter and Yogev in the \clique model, on constructing graph spanners in unweighted graphs~\cite{parter2018spanner} of $O(k)$ stretch.
In \cref{sec:O(k)-spanners}, we show how by building on this work one can obtain the following result.
\begin{theorem}\label{thm:super-linear-memory}
Given an unweighted graph $G$ on $n$ vertices and $m$ edges, there is an algorithm that in $O(\frac{1}{\gamma} \cdot \log k)$ $\MPC$ rounds w.h.p.~outputs a spanner of size $O(n^{1+1/k} \cdot k)$ and stretch $O(k)$ when memory per machine is $O(n^\gamma)$, for any constant $\gamma > 0$. This algorithm uses a total memory of $\tO(m + n^{1 + \gamma})$.
\end{theorem}



It is worth noting that these round complexities are close to optimal, conditioned on a widely believed conjecture. This is because lower bounds in the distributed LOCAL model imply an $\Omega(\log k)$ conditional lower bound for the closely related problem of finding spanners with optimal parameters, i.e., $(2k-1)$-spanners with $O(n^{1+\frac{1}{k}})$ edges.
Specifically, there is an $\Omega(k)$ lower bound for the problem in the LOCAL model \cite{derbel2008locality}, and as shown in \cite{GhaffariKU19}, this implies $\Omega(\log k)$ conditional lower bound in $\MPC$ with sublinear memory, under the widely believed conjecture that connectivity requires $\Omega(\log n)$ rounds.\footnote{The proof in \cite{GhaffariKU19} is for $\MPC$ algorithms that are component-stable, see \cite{GhaffariKU19} for the exact details.} This conjecture is also called the cycle vs two cycles conjecture, as even distinguishing between one cycle of $n$ nodes from 2 cycles of $n/2$ nodes, is conjectured to require $\Omega(\log n)$ rounds. 
Our algorithms have complexity of $\poly(\log k)$ for near-optimal (up to a factor of $k^{o(1)}$) stretch which nearly matches the lower bound.

It is an interesting question whether spanners with \emph{optimal} parameters can be also constructed in $\poly(\log k)$ time. While our main goal is to construct spanners in standard $\MPC$ in $\poly \log k$ rounds, we also provide two results that lead to near optimal parameters (stretch $O(k)$ and size $\tilde{O}(n^{1+1/k})$) at the expense of computational resources. 
As mentioned above, \cref{thm:super-linear-memory} constructs $O(k)$-spanners with $O(n^{1+1/k} \cdot k)$ edges in $O(\log{k})$ rounds. This however comes at a price of extra total memory, and also only works for \emph{unweighted} graphs.
Additionally, in \cref{sec:polyround_algorithm}, we provide an algorithm that runs in $O(\sqrt{k})$ rounds and computes an $O(k)$-spanner with $O(\sqrt{k} n^{1+1/k})$ edges. While the parameters of the spanner are near-optimal, this comes at a price of significant increase in the round complexity (this is still much faster compared to previous algorithms that require $O(k)$ time).

\paragraph{Distance Approximations.}
As a direct but important corollary  of our spanner construction in \Cref{corollary:k-log 3}, in the near-linear memory regime we get an $O(\log^2\log n)$-round algorithm for $O(\log^{1+o(1)} n)$ approximation of distance related problems, including all-pairs-shortest-paths. Alternatively, we can obtain a faster algorithm that runs in $O(\log\log n)$ rounds, if we relax the approximation factor to $O(\log^{1+\eps} n)$ for a constant $\eps>0$. 
It is important to note that as the global memory in MPC is bounded by $\tilde{O}(m)$, we do not have enough space to store the complete output of APSP. Instead, we have one \textit{coordinator} machine that stores the spanner, which implicitly stores approximate distances between all vertices. This machine can then compute distances locally based on the spanner.
This gives the following (See \cref{sec:APSP} for full details).

\begin{restatable}{corollary}{appAPSP}
There is a randomized algorithm that w.h.p computes $O(\log^s n)$-approximation for APSP in weighted undirected graphs, and runs in $O(\frac{t\log \log{n}}{\log(t+1)})$ rounds of $\MPC$, when memory per machine is $\tO(n)$, $s= \frac{\log (2t+1)}{\log(t+1)}$, and $t=O(\log{\log{n}})$. This algorithm uses total memory of size $\tO(m)$.
\end{restatable}

Our work leaves an intriguing open question. 

\medskip
\noindent\textbf{Open Problem.} \; \emph{Can we compute a constant, or perhaps even $1+o(1)$, approximation of all pairs shortest paths in $\poly(\log\log n)$ rounds in the near-linear local memory regime of $\MPC$?}

\paragraph{Extension to Other Models.}
In addition to these two models, we show the generality of our techniques by extending our results to the PRAM and \clique models. 
\subparagraph{\clique} We can extend our spanner construction to work also in the distributed \clique model (see \cref{sec:clique}), leading to the following corollary for approximate shortest paths.

\begin{restatable}{corollary}{APSPclique}
There is a randomized algorithm in the \clique model that w.h.p computes $O(\log^s n)$-approximation for APSP in weighted undirected graphs, and runs in $O(\frac{t\log \log{n}}{\log(t+1)})$ rounds, where $s= \frac{\log (2t+1)}{\log(t+1)}$, and $t=O(\log{\log{n}})$.
\end{restatable}

As two special cases, this gives $O((\log n)^{\log 3})$-approximation in $O(\log \log n)$ time, and $O(\log^{1+o(1)}n)$-approximation in $O((\log{\log n})^2)$ time.
It is important to note that these are the first \emph{sub-logarithmic} algorithms that approximate weighted APSP in the \clique model. Prior results take at least poly-logarithmic number of rounds or only work for \emph{unweighted} graphs, as we review next. 

Previous poly-logarithmic algorithms include the following. First, the spanner construction of Baswana-Sen \cite{baswana_sen} gives $O(\log n)$-approximation for weighted APSP in $O(\log n)$ rounds, by building $O(\log{n})$-spanner of size $O(n\log{n})$ and letting all vertices learn the spanner. Additionally, it is possible to get a $(3+\epsilon)$-approximation for weighted APSP in $O(\frac{\log^2{n}}{\epsilon})$ time \cite{censor2019fast}. In unweighted graphs, a recent work \cite{CliqueShortestPaths2020} shows that it is possible to get $(2+\epsilon)$-approximation for APSP in $poly(\log{\log{n}})$ time, however the techniques only apply for the unweighted case.

We note that while there are fast constructions of spanners in the \clique model \cite{parter2018spanner}, they work only for \emph{unweighted} graphs, and also the size of the spanner is at least $\Omega(n \log n)$. The latter adds $\Omega(\log n)$ rounds to the complexity of APSP, for collecting the spanner, which means one cannot achieve a sublogarithmic round complexity for APSP using these spanner constructins.

\subparagraph{PRAM}
Our result also extends to the PRAM CRCW model. We would get the same PRAM depth as the MPC round complexity, with an additional multiplicative $\log^* n$ factor
that arises from certain PRAM primitives (see \cref{app:implementation}).
We note that $O(k \log^{*}n)$ depth PRAM algorithms for $O(k)$ spanners were studied in \cite{miller2015improved}, and \cite{baswana_sen}. To the best of our knowledge, before this work there were no known algorithms in PRAM with depth $o(k)$, even for suboptimal spanners.

\section{Overview of Our Techniques}
\label{sec:overview}


The main objective of our work is to design spanners that can be implemented in small parallel depth, e.g., $\poly(\log\log n)$ rounds of MPC. For the sake of simplicity, we present our main ideas by considering unweighted graphs only.
However, our main results apply to weighted graphs as well, and that's one of the key strengths of our approach.

Our general approach will be to grow clusters of vertices by engulfing adjacent vertices and clusters.
Additionally, each cluster will be associated with a rooted tree that essentially shows the ``history'' of the cluster's growth,
with the root representing the oldest vertex of the cluster.
The tree edges will all be part of the spanner, and the maximum depth of the tree will be called the \emph{radius} of the corresponding cluster.

A general paradigm when attaching a cluster to either a vertex or another cluster will be to keep only one edge between them
(the one of minimum weight) in the spanner, and discarding the remaining edges.
The discarded edges will be spanned by the single included edge and the tree edges within the cluster(s).
Thus, (disregarding many details) we can generally upper bound the stretch of any discarded edge by four times the maximum radius of any cluster.
Moving forwards, we will talk about stretch and cluster radius interchangeably since they are asymptotically equal.

\subsection{Fast Algorithm using Cluster-Cluster Merging}
\label{sec:a_fast_algorithm_with_poor_stretch}
As a starting point, we design an algorithm based on \emph{cluster-cluster merging},
that can be summarized as follows.

\medskip
\smallskip
\begin{minipage}{0.95\linewidth}
\begin{mdframed}[hidealllines=true, backgroundcolor=gray!15]
\vspace{-3pt}
\textbf{Cluster-cluster merging}\\
Given an unweighted $G = (V, E)$ and parameter $k \geq 2$:
\begin{itemize}
\item[-] Let $\cC_0$ be a set of $n$ clusters ($|V|=n$), where each vertex of $V$ is a single cluster.
\item[-] For $i = 1 \ldots \log{k}$:
    \begin{enumerate}[(A)]
        \item\label{item:sample-clusters-cluster-merging} Let $\cC_i$ be a set of clusters obtained by sampling each cluster from $\cC_{i - 1}$  with probability $n^{(-2^{i - 1} / k)}$. 
        \item\label{item:each-cluster-cluster-merging} For each cluster $c \in \cC_{i - 1} \setminus \cC_i$:
        \begin{enumerate}[(i)]
           \item\label{step:merge-cluster-merging} If $c$ has a neighbor $c' \in \cC_i$, add an edge between $c$ and $c'$ to the spanner, and merge $c$ to $c'$.
            \item\label{step:add-edge-unsampled-cluster-merging} Otherwise, for every $c' \in \cC_{i - 1} \setminus \cC_i$ that is a neighbor of $c$ add an edge between $c$ and $c'$ to the spanner.
        \end{enumerate}
    \end{enumerate}
\item[-] Add an edge between each pair of clusters remaining.
\end{itemize}
\end{mdframed}
\end{minipage}
\medskip

We start by letting each vertex be a singleton cluster, and in each subsequent iteration, a set of clusters is sub-sampled.
These \emph{sampled clusters} are ``promoted'' to the next iteration (the remaining clusters will not be considered in future iterations).
However, the \emph{sampled} clusters grow to absorb the neighboring \emph{unsampled} ones.
Conversely, each unsampled cluster joins its nearest neighboring sampled cluster.
If no such neighboring (sampled) cluster exists, then we can conclude that, with high probability, there are very few neighboring clusters,
and this allows us to deal with these problematic clusters by including edges to all their neighbors in the spanner.

\subparagraph*{Intuition for the Analysis.}
The intuition behind this approach is as follows.
As in each iteration clusters are merged, the number of clusters decreases significantly between iterations.
Since we are aiming for a spanner of size $\tilde{O}(n^{1+1/k})$, if $C$ is the current set of clusters, we can allow each cluster in $C$ to add $\frac{n^{1+1/k}}{|C|}$ adjacent edges to the spanner.
This means that when the number of clusters reduces, we can allow each cluster to add significantly more edges to the spanner.
This allows us to decrease the sampling probability $p$ between different iterations drastically. As a consequence, the number of clusters decreases rapidly, which results in a $\log{k}$ complexity.
We show that in each iteration the merging grows the stretch by a factor of 3, leading to an overall stretch of $O(3^{\log{k}})=O(k^{\log{3}}).$ 


\subsection{Obtaining Better Stretch}
\label{sec:obtaining_better_stretch}

We now discuss how to reduce the stretch, all the way to $k^{1 + o(1)}$,
while maintaining the same spanner size and increasing the number of iterations only by an extra $\log{k}$ factor.

Intuitively, the spanner algorithm described has sub-optimal stretch of $k^{\log{3}}$ mainly since it is too aggressive in growing clusters in each iteration, and when two clusters are merged, the radius increases by a factor of three in one iteration. Thus a natural idea would be to grow the clusters much more gradually;
that is, instead of merging a cluster $c$ to cluster $c'$ entirely in one iteration,
$c'$ could consume parts of $c$ repeatedly (over multiple iterations). In other words, similar to the algorithm of \cite{baswana_sen} we can grow the clusters incrementally before performing a merge. We call this process \emph{cluster-vertex merging}.

\subparagraph{Cluster-vertex merging}
The main difference between cluster-cluster and cluster-vertex merging can be outlined as follows:
Firstly, in \cref{item:sample-clusters-cluster-merging} use a smaller sampling probability, e.g. $n^{-1/k}$, instead of $n^{-2^{i - 1} / k}$;
Secondly, in \cref{step:merge-cluster-merging} instead of merging two clusters, merge to a sampled cluster $c'$,
only the vertices \emph{incident} to $c'$ that \emph{do not belong} to any sampled cluster.
This essentially recovers the algorithm in \cite{baswana_sen}, where it was shown that using only the cluster-vertex merging,
one can construct a spanner of stretch $O(k)$, but this also requires $O(k)$ rounds.
We investigate the interpolation between these two extremes by using a hybrid approach,
which uses both incremental cluster growing (cluster-vertex merging) and cluster-cluster merging.

\subparagraph{Combining the two approaches using cluster contractions.}
The basic idea is as follows.
Instead of merging clusters in each iteration,
we alternate between iterations where we apply the cluster-vertex merging approach and iterations where we apply the cluster-cluster merging approach.
After a few iterations where we apply cluster-vertex merging, we merge clusters (an operation we also refer to as \emph{contraction}).
this allows to get an improved stretch but still keep a small running time.

Running the \emph{cluster-vertex merging} procedure on a contracted graph also results in the cluster size growing much faster in each step.
This accelerated cluster growth is the main reason for the speedup in our algorithm.

Intuitively, contractions result in \emph{loss of information} about the \emph{internal structure} of the cluster,
and any time they are performed we incur extra stretch penalties.
Our aim is to carefully balance out this loss with the cluster growth rate to get the best tradeoffs. We adjust this by tuning the sampling probabilities, and the intervals at which we perform contractions. This interpolation will allow us to reduce the stretch to $k^{1 + o(1)}$, while still having $O(\poly \log k)$ round (iteration) complexity.  More generally, our various tradeoffs are a consequence of how much we grow the clusters before each contraction.

\remove{
\subsubsection{bsection name}
\label{sec:bsection_name}

Before answering this question, we provide some intuition on different parts of cluster-merging algorithm. First, it is useful to think of each cluster of $\cC_i$ being contracted to a vertex. In other words, we associate a graph $G_i$ to set of clusters $\cC_i$, where a cluster of $\cC_i$ corresponds to a single vertex in $G_i$. (In fact, in the subsequent section we will reason about clusters in this contracted way.) This means that from iteration to iteration the number of vertices in $G_i$ reduces at increased rate, implying that after only a couple of steps $G_i$ is significantly smaller than $G$. Hence, an important ramification is that the number of vertices in the contracted graph $G_i$ can be much smaller than the original – specifically, it depends on the previous sub-sampling steps and the corresponding probabilities.

Second, as a thought experiment, imagine that the sampling probability in \cref{item:sample-clusters-cluster-merging} of cluster-merging equals $0$. In this case, no cluster would be sampled in $\cC_1$, and hence for each cluster of $\cC_0$  \cref{step:add-edge-unsampled-cluster-merging} would be executed, resulting in all edges being added to the spanner. On one hand, the output spanner would have stretch $1$, but on the other hand its size would be the same as the input graph (which is too large). 
A takeaway message from this, which will be true for all our algorithms, is that decreasing sampling probability reduces the number of iterations and in turn leads to better stretch.

Now, our main approach can be summarized as follows: execute a few (less than $\log{k}$) iterations of cluster-merging with the goal to reduce the graph size; after the graph size is reduced, perform a number of cluster-growth steps (significantly fewer than $k$) with aggressive sampling probability; repeat. Running cluster-growth procedure on a graph $G_i$ of contracted clusters also results in the cluster size growing much faster in each step
(engulfing a single super node is the same as engulfing an entire cluster).
This accelerated cluster growth is the main reason for the speedup in our algorithm. On the other hand, merges result in \emph{loss of information} about the \emph{internal structure} of the cluster,
and thus, we incur extra stretch penalties.
Our aim is to carefully balance out this loss with the cluster growth rate to get the best tradeoffs.
We adjust this by tuning the sampling probabilities, and the intervals at which we perform cluster-merging.
Our various tradeoffs are hence a consequence of how much we grow a cluster before each merge.
}

\subsection{The General Algorithm for Round-Stretch Tradeoffs}
\label{sec:incorporating_cluster_contraction_to_obtain_new_algorithms}

Now, we provide a more detailed overview of our general algorithm.
We will then look at specific parameter settings that lead to various tradeoffs. 
The algorithm proceeds in a sequence of \textit{epochs}, where epoch $i$ consists of $t$ iterations as follows:
In each iteration we subsample the clusters with probability $p_i$ (to be defined later).
As before, we then grow each sampled cluster by adding all neighboring vertices that have not joined any other sampled cluster.
At this point we contract the clusters, and move to the next epoch.

Our general algorithms (in \cref{sec:general_algorithm}) achieve a range of tradeoffs,
parameterized by $t$, which is the number of growth iterations before a contraction, as we outline below.
\begin{enumerate}[(A)]
    \item Assume that the number of \emph{super-nodes} in the current graph is $n'$ (originally $n$).
    \item Repeat $t$ times (essentially performing $t$ iterations of \cite{baswana_sen} with adjusted sampling probabilities):
    \begin{enumerate}[1.]
        \item Perform \emph{cluster-sub-sampling} with $p_i = \frac{n'}{n^{1+1/k}}$:\\
        Ensures that number of added edges $\frac{n'}{p_i} = O(n^{1+\frac{1}{k}})$.
        \item Perform incremental \emph{cluster-vertex merging} to increase the radius of each cluster by one unit in the \emph{current graph}:\\
        Note that the current graph may contain contracted clusters as \emph{super-nodes}.
        Assuming that the internal radius of the \emph{super-nodes} is $r$,
        the actual radius increase in the original graph can be as large as $2r+1$ units (see \cref{fig:loground_strong_radius_induction}).
    \end{enumerate}
    \item Perform \emph{cluster-contraction} on the most recent clusters (clusters become super-nodes).\\
    This has the effect of reducing the number of clusters in the graph by a factor of $\pc^{t}$
    (probability of a specific cluster surviving for $t$ repititions of \emph{cluster-sub-sample}).
    So, we get $n' \leftarrow n'\cdot \pc^t$.
\end{enumerate}

Following are some example results we obtain for specific values of parameter $t$.
\begin{itemize}
    \item $\mathbf{(t=k)}$:
    This is one extreme case where there is \emph{no contraction}, and the cluster radius only increases by $\mathsf 1$ for $k$ repetitions,
    thus recovering the \cite{baswana_sen} result.
    This is the \emph{slowest} algorithm, but it achieves optimal stretch $2k-1$.
    \item $\mathbf{(t=\sqrt{k})}$:
    The immediate generalization involves exactly one contraction (analysis in \cref{sec:polyround_algorithm}),
    which occurs after the first set of $\sqrt{k}$ iterations.
    A new sampling probability is introduced after the contraction (reduced size of graph).
    According to this probability, the remainder of the algorithm is actually just a normal $\sqrt{k}$ stretch spanner construction.
    This algorithm attains $O(k)$ stretch, but it dramatically reduces the number of rounds to $O(\sqrt{k})$.
    \item $\mathbf{(t=1)}$:
    This is the other extreme case (analyzed in \cref{sec:loground_cluster_merging}),
    where the three procedures (sub-sample, grow, contract) are performed repeatedly one after another, i.e., the algorithm contracts immediately after a single grow step. Consequently, the cluster radius grows exponentially, and the algorithm terminates after $\log k$ repetitions,
    yielding Item~\ref{item:one_BS_step} of \cref{corollary:k-log 3}.
    This is the \emph{fastest} algorithm, but only achieves stretch $O(k^{\log 3})$.
\end{itemize}

One special interesting setting, which we also use for application in distance approximation, is when we set $t= \log k$. This leads to stretch $k^{1+o(1)}$ and requires only $O(\frac {\log^2 k}{\log\log k})$ rounds. The general tradeoffs can be found in \cref{sec:general_algorithm}.

\subsection{Related Spanner Constructions} 

Our approach can also be seen as a new contraction-based spanner algorithm with a focus on parallel depth/round  efficiency.

In the context of dynamic stream algorithms, another contraction-based algorithm was proposed by \cite{AhnGM12}, but only for unweighted graphs. Their contractions are based on a different type of clustering formed based on vertex degrees. Algorithm of \cite{AhnGM12} has resemblance to a special case of our algorithm described in \cref{sec:loground_algorithm}, but it has a weaker stretch. In particular, for a spanner of size $\tilde{O}(n^{1+1/k})$ they obtain stretch $k^{\log 5}$ in $\log k$ passes in streaming (where a pass corresponds to one round of communication in \MPC) in \textit{unweighted graphs}, whereas in the same time (pass/round) we obtain stretch $k^{\log 3}$ even for \textit{weighted graphs}. Our general algorithm, as a special case, obtains the much stronger stretch of $k^{1+o(1)}$ in $O(\frac{\log^2 k}{\log \log k})$ iterations. 

These contraction-based algorithms can be seen as an alternative approach to the well-known algorithm of \cite{baswana_sen}. Mainly, the goal in \cite{baswana_sen} is to compute optimal spanners of stretch $2k-1$, whereas our main goal is time efficiency. As a result our algorithm has a slightly weaker stretch/size tradeoff, but requires exponentially fewer iterations.

This general contraction-based framework may be of interest also in other related distance objects. A related work by \cite{ben2020new} focuses on $(\alpha,\beta)$-spanners and hopsets, and they use a similar type of clustering as one part of their construction. However, they connect the clusters differently and in since their main focus is not computation time, their algorithms run in polynomial time in most models. We hope that our fast clustering techniques also give insight into faster algorithms for these structures, perhaps at an extra cost in the stretch. Such improvements will have immediate implications for distance computation in various models. 


After the submission of our initial manuscript in March 2020, we found out  that an independent and concurrent work \cite{filtser2021} obtained similar bounds for spanner construction in dynamic stream settings. Concretely, \cite{filtser2021} (see Theorem 3, and set $g=\epsilon \cdot \log k$), leads to a streaming algorithm with $O(\epsilon \cdot \log k \cdot 2^{1/\epsilon})$ passes, spanner stretch $O(k^{1+\epsilon})$ and size $\tilde{O}(n^{1+1/k})$. This matches our MPC bound (replacing the number of passes by number of rounds), by setting $t= 2^{1/\epsilon}$ in Theorem \ref{theorem:main}. For weighted graphs, they can obtain a spanner with an extra factor of $\log W$ in the size ($W$ is the aspect ratio), whereas in our construction the size is the same for weighted and unweighted graphs.

\section{Cluster-Contraction Algorithm for Near-Optimal Spanners}
\label{sec:polyround_algorithm}


As a warm-up, in this section we discuss an algorithm that takes $O(\sqrt{k})$ rounds and constructs a spanner with stretch $O(k)$ and size $O(\sqrt{k}n^{1+1/k})$ in unweighted graphs. Note that this simple algorithm is already significantly faster compared to \cite{baswana_sen}, requiring only $O(\sqrt{k})$ rounds instead of $O(k)$ rounds. 

Given an unweighted graph $G=(V,E)$ we compute a spanner of size $O(\sqrt{k}n^{1+1/k})$ with stretch $O(k)$. The high-level idea is as follows. We run the algorithm of \cite{baswana_sen} \textit{twice}: in the first phase we perform the first $t=\Theta(\sqrt{k})$ iterations of \cite{baswana_sen} and stop. We will form a supergraph $\hat{G}=(\hat{V}, \hat{E})$ defined by setting each cluster $\cC_t$ to be a supernode, and will add an edge between supernodes $c_1, c_2 \in \hat{V}$ if the corresponding clusters are connected with at least one edge in the original graph $G$. Now, for $t' = \Theta(\sqrt{k})$, we compute a $t'$-spanner on $G$ by running the \cite{baswana_sen} algorithm on the graph $\hat{G}$ as a black-box, this requires only $t' = \sqrt{n}$ additional iterations. 
Next we describe these two phases in detail.

\paragraph{First Phase:}
Start with $\cR_0 = V$, and $V'=V$, $\rE=E$. We will have a sequence of clusterings $\cR_1 \supseteq \ldots \supseteq \cR_{t}$ for a parameter $t$ (we will set $t=\sqrt{n}$). $V'$ and $\rE$ will be the set of vertices and edges that are not yet \textit{settled}. 
In each iteration of the first phase, $V'$ is the set of vertices with one endpoint in $\rE$. 
\begin{enumerate}
\item Sample a set of clusters $\cR_i$ by choosing each cluster in $\cR_{i-1}$ with probability $n^{-1/k}$. Set $\cC_i = \cR_i$. 
\item For all $v \in V'$:
\begin{enumerate}[(i)]
\item If $v$ is adjacent to a sampled cluster $\cR_i$, then add $v$ to the closest $c \in \cC_i$ and add one edge from $E(v,c)$ to the spanner. Discard (remove from $\rE$) all the edges in $E(v,c)$.

\item If $v$ is not adjacent to any sampled clusters in $\cR_i$, then for \textit{each} neighboring cluster $c' \in \cC_{i-1}$ add a single edge from $E(v,c')$ to the spanner, and discard all other edges between $E(v,c')$. 
\item Remove the intra-cluster edges: remove all the edges with both endpoints in $\cC_i$ from $\rE$.
\end{enumerate}

\end{enumerate}

\paragraph{Second Phase:}
Define a supergraph $\hat{G}=(\hat{V},\hat{E})$ by setting each cluster $\cC_t$ to be a node in $\hat{G}$ and adding an edge in $\hat{E}$ for each pair of adjacent clusters in $\cC_t$. We then run a black-box algorithm for computing a $(2t'-1)$-spanner (e.g. by running the algorithm of \cite{baswana_sen}) on $\hat{G}$, for $t' = \sqrt{n}$.


\paragraph{Analysis Sketch.} The high-level idea is that we are stopping the algorithm of \cite{baswana_sen} when there are $O(n^{1-1/\sqrt{k}})$ (or more generally $O(n^{1-t/k})$) clusters in $C_t$. This means that the supergraph $\hat{G}$ is significantly smaller, and now we can afford to compute a spanner with a better stretch on $\hat{G}$, for fixed size. The radius of the clusters at termination is $O(tt')=O(k)$, and thus the overall stretch is $O(k)$. To formalize this argument, we start with the size analysis.

\begin{theorem}
The set of edges added by this algorithm is $O(\sqrt{k}n^{1+1/k})$.
\end{theorem}
\begin{proof}[Proof.]
In the first phase we only add as many edges as the \cite{baswana_sen} algorithm does and the total number of edges is $O(tn^{1+1/k})$. In the second phase we will add $O(n^{1-t/k})^{1+1/t'})$ edges, which is $O(n^{1-1/k})$ for $t=t'= \sqrt{n}$.
\end{proof}

We provide a brief argument for the stretch. A similar and more formal argument for similar claims will be presented in \cref{sec:loground_cluster_merging}.

We next provide the stretch analysis of the algorithm.
We first argue that if an edge $(u,v)$ is not added to the spanner, or discarded by the end of iteration $i$, then the endpoints $u$ and $v$ are in distinct clusters in $\mathcal{C}_i$. This would imply that we can only restrict our attention to the vertices that survive inside clusters. Otherwise, they are removed from $V'$. 
\begin{restatable}{lemma}{polyroundRemainingEdges}
\label{lem:polyround_remaining_edges}
Let $(u,v) \in E$ be an edge not added to the spanner.
At the end of iteration $i$ of the first phase, $(u,v)$ is either discarded (removed from $\rE$),
or both its endpoints belong to clusters in $\cC_i$.
\end{restatable}
\begin{proof}
Let $j \leq t$ be the first iteration in which at least one of $u$ or $v$ does not belong to $\cC_j$. In other words, their cluster does not get \textit{promoted} (i.e. is not a cluster subsampled at level $j$). W.l.o.g assume that $u$ is the endpoint that is not part of any cluster in $\cC_j$. In other words, $u$ does not have any neighbouring sampled cluster in $\cC_j$. By our assumption on $j$, we know that there exists a cluster $c_v \in \cC_{j-1}$ that $v$ belongs to. Hence if edge $(u,v)$ is not added to the spanner, it must have been discarded in step 2 (ii). 
\end{proof}

We will also use the following property of the algorithm as shown in \cite{baswana_sen}. 

\begin{restatable}{lemma}{basicRadius}
\label{lem:basic_radius}
At iteration $i$ of the first phase, all clusters $\cC_i$ have radius $i$.
\end{restatable}

We now use the properties described to prove the overall stretch.
\begin{theorem}
For each edge $(u,v) \in E$ not added to the spanner there is a path of length $O(k)$ between $u$ and $v$.
\end{theorem}
\begin{proof}


Let $(u,v) \in E$.
From \cref{lem:polyround_remaining_edges}, by the end of the first phase, $(u,v)$ is either discarded, or its endpoints belong to clusters in $\cC_t$.
We start by discussing the case it is discarded, we consider the following two cases.

\emph{Case 1: the edge $(u,v)$ is removed in step 2 (i) or (ii).}  Assume w.l.o.g. that this edge was removed while processing node $u$. In this case, an edge $(u,v')$ was added to the cluster $c_v \in \cC_i$ that $v$ belongs to. Given the fact that this cluster has radius $i$, we know that there is a path of length $2i+1$ between $u$ and $v$.

\emph{Case 2: $(u,v)$ is removed in step (iii).} Here $u$ and $v$ both are assigned to the same cluster, say $c \in \cC_i$. This means (by \cref{lem:basic_radius}) that there is a path of length at most $i$ from $u$ to center of $c$, and a path of length at most $i$ from $v$ to center of $c$. Thus there is a path of length at most $2i$ between $u$ and $v$. Hence at the end of first phase all clusters have radius $2t$. 

We next consider the case that by the end of the first phase, both of the endpoints of $(u,v)$ belong to clusters in $\cC_t$.  If they belong to the same cluster by iteration $t$, then we know that there is a path of length $2t$ between $u$ and $v$, since this cluster has radius at most $t$. Otherwise $u$ and $v$ belong to distinct clusters in $\cC_t$, which we denote by $c_u$ and $c_v$ respectively. These clusters will be ``supernodes" in $\hat{V}$. In other words, any vertex that is not part of clusters in $\cC_t$ is already processed and removed from $V'$. Therefore we can restrict our attention to vertices that are part of clusters in $\hat{V}$. 

Since we construct a $t'$-spanner on $\hat{G}$, at the end of the second phase, there will be a path of length $t'$ in $\hat{G}$ between $c_v$ and $c_u$. Since each of the clusters in $\hat{G}$ have radius $t$ at the time they get passed to the second phase, a path on $\hat{G}$ with stretch $O(t')$ corresponds to a path in $G$ with stretch $O(tt')$. The claim follows when $t=t'=\sqrt{n}$.
\end{proof}

\remove{
\paragraph{Stretch Analysis Sketch.}
The stretch analysis appears in \cref{sec4_appendix}.
We next give a high-level overview of it.
First, we show the following.

\begin{restatable}{lemma}{polyroundRemainingEdges}
\label{lem:polyround_remaining_edges}
Let $(u,v) \in E$ be an edge not added to the spanner.
At the end of iteration $i$ of the first phase, $(u,v)$ is either discarded (removed from $\rE$),
or both its endpoints belong to clusters in $\cC_i$.
\end{restatable}

\begin{restatable}{lemma}{basicRadius}
\label{lem:basic_radius}
At iteration $i$ of the first phase, all clusters $\cC_i$ have radius $i$.
\end{restatable}

We use the properties described to prove that the stretch is $O(k)$.
Intuitively, if we take an edge $(u,v) \in E$, by \cref{lem:polyround_remaining_edges}, by the end of the first phase this edge is either discarded, or its endpoints belong to clusters in $\cC_t$. If it was discarded during the first phase, it follows that there is a path of stretch $O(t)=O(\sqrt{k})$ between $u$ and $v$ based on the analysis of \cite{baswana_sen}. If  $(u,v)$ survived the first phase, its endpoints are either in the same cluster in $\cC_t$ which implies a path of stretch $O(t)$ between them by \cref{lem:basic_radius}, or they belong to different super-nodes in $\hat{V}$. In the latter case, since in the second phase we compute an $O(t')$-spanner in $\hat{G}$, there is a path of stretch $O(t')=O(\sqrt{k})$ between the  super-nodes corresponding to $u$ and $v$, which implies a stretch of $O(tt')=O(k)$ between $u$ and $v$.
See \cref{sec4_appendix} for details.
}

\section{Cluster-merging Approach}
\label{sec:loground_cluster_merging}

Following the general paradigm of Baswana and Sen~\cite{baswana_sen}, our algorithm proceeds in two phases. 
In the first phase, the algorithm creates a sequence of growing clusters, where initial clusters are singleton vertices.
This is similar to the first phase in the \cite{baswana_sen} algorithm, but has the following crucial differences:
\begin{itemize}
    \item In each epoch, a sub-sampled set of clusters (from the previous epoch) expand,
        by engulfing \emph{neighboring clusters} that were \emph{not} sub-sampled.
        In \cite{baswana_sen}, the sub-sampled clusters only engulf \emph{neighboring vertices}.
    \item Consequently, the radius of our clusters increase by a factor of $3$ (roughly) in every epoch,
        and thus the radius after epoch $i$ is $O(3^i)$.
        On the other hand, the cluster radius in \cite{baswana_sen} increments by $1$ in each epoch,
        leading to a radius of $i$ at the end of epoch $i$.
    \item The sub-sampling probability at epoch $i$ is $n^{-\frac{2^{i-1}}{k}}$,
        i.e., the probabilities \emph{decrease} as a \emph{double exponential},
        as opposed to \cite{baswana_sen}, where the probabilities are always the same.
    \item Our algorithm proceeds for $O(\log k)$ epochs
        as opposed to $k$ epochs in \cite{baswana_sen}.
    \item The final stretch we achieve is $O(k^{\log 3})$, instead of $O(k)$ as in \cite{baswana_sen}.
\end{itemize}



Based on the aforementioned sampling probabilities, the final number of clusters will be $n^{1/k}$.
Subsequently, we enter the second phase, where we add edges between vertices that still have un-processed edges and the final clusters.
The final output is a set of edges $E_S$ that represent the spanner.

The main benefit of our approach, compared to \cite{baswana_sen}, is that it provides a significantly faster way of constructing spanners in $\MPC$. Namely, the algorithm of \cite{baswana_sen} inherently requires $O(k)$ iterations and it is not clear how to implement it in $o(k)$ $\MPC$ rounds while not exceeding a total memory of $\tO(m)$. On the other hand, in \cref{app:implementation} we show that each epoch of our algorithm can be implemented in $O(1)$ $\MPC$ rounds. This implies the above approach can be implemented in $O(\log{k})$ $\MPC$ rounds.

\subsection{Algorithm}
\label{sec:loground_algorithm}
In this section we describe the cluster-merging approach.
We will use the following notation for inter-cluster edges.
\begin{definition}
\label{def:inter_cluster_edges}
We define $\rE(c_1, c_2)$ to be the set of edges in $\rE$ that have one endpoint in cluster $c_1$ and the other endpoint in cluster $c_2$.
We will also abuse this notation, and use $\rE(v, c)$ to denote all edges between \emph{vertex} $v$ and cluster $c$.
\end{definition}
Before a formal description of our algorithm, we also need two more definitions.
\begin{definition}
\label{def:cluster}
A \emph{cluster} $c$ is a set of vertices $V_c\in V$ along with a rooted tree $T(c) = (V_c, E_c)$.
The ``center'' of the cluster is defined as the root of $T$ (the oldest member),
and the ``radius'' of the cluster is the depth of $T$ (from the root).
\end{definition}

\begin{definition}
\label{def:clustering}
A clustering of a graph $G=(V,E)$ is a partition of $V$ into a set of disjoint clusters $\cC$, such that for all $c, c'\in \cC$,
we have $V_c\cap V_{c'} = \emptyset$.
\end{definition}

We next describe the two phases of our algorithm.
\subparagraph{Phase 1:}
The first phase proceeds through $\log k$ epochs.
Let $\cC_0$ be the clustering of $V$ where each $v \in V$ is a cluster.
At a high-level, during epoch $i$, we sub-sample a set of clusters $\cR^{(i)}\subseteq \cC^{(i-1)}$,
we will connect the clusters to each other as follows:
Consider a cluster $c$ that is \emph{not} sampled at epoch $i$.
If $c$ does \emph{not} have a \emph{neighboring sampled cluster} in $\cR^{(i)}$,
we merge it with \textit{each} of the (un-sampled) neighboring clusters in $\cC^{(i-1)}$, using the lowest weight edge.
On the other hand, if $c$ has \emph{at least one} neighboring sampled cluster,
then we find the closest such cluster and merge it to $c$ using the lowest weight edge, say $e$.
Additionally, in the weighted case, we also add an edge to each of the other neighboring clusters,
that are adjacent to $c$ with an edge of \emph{weight strictly lower than} that of $e$.

Throughout, we maintain a set $\rE$ (initialized to $E$) containing the unprocessed edges.
During each epoch, edges are removed from $\rE$.
During execution, some edges from $\rE$ are added to the set of spanner edges $E_S$, and some are discarded.
Specifically, when we merge clusters $c_1$ and $c_2$, only the lowest weight edge in $\rE(c_1, c_2)$ is added to the spanner $E_S$,
and all other edges in $\rE(c_1, c_2)$ are discarded from $\rE$ (the notation $\rE(\cdot, \cdot)$ is in \cref{def:inter_cluster_edges}).
We will use $\rE^{(i)}$ to denote the state of the set $\rE$ at the end of epoch $i$.
During epoch $i$, we also construct a set of edges $\cE^{(i)} \subseteq E_S$,
containing the edges that are used to connect (merge) \emph{sampled} old clusters with \emph{un-sampled} clusters to form the new ones $\cC^{(i)}$.

At epoch $i$, for $i = 1 \ldots \log{k}$, we perform the following steps:
\begin{enumerate}[(1)]
    \item\label{item:loground_phase-1-step-1} Sample a set of clusters $\cR^{(i)} \subseteq \cC^{(i-1)}$,
        where each $c\in \cC^{(i-1)}$ is chosen to be a member of $\cR^{(i)}$ with probability $n^{-\frac{2^{i-1}}{k}}$.
        Initialize $\cE^{(i)}$ to the subset of edges in $\cE^{(i-1)}$ that are contained in some cluster $c\in \cR^{(i)}$.

    \item\label{item:loground_phase-1-step-2} Consider a cluster $c\in \cC^{(i-1)} \setminus \cR^{(i)}$ that has a neighbor in $R^{(i)}$.
        Let $\cN^{(i)}(c)\in \cR^{(i)}$ be the \emph{closest neighboring sampled cluster} of $c$.

    \begin{enumerate}[(a)]
        \item\label{item:loground_phase-1-step-2-step-i} Add the lowest weight edge $e \in \rE(c, \cN^{(i)}(c))$
            to both $\cE^{(i)}$ and $E_S$ and remove the entire set $\rE(c, \cN^{(i)}(c))$ from $\rE$.
        \item\label{item:loground_phase-1-step-2-step-ii}
            For all clusters $c'\in \cC^{(i-1)}$ adjacent to $c$ with any edge of weight strictly less than $e$,
            add the lowest weight edge in $\rE(c, c')$ to $E_S$ and then discard all the edges in $\rE(c, c')$ from $\rE$.
    \end{enumerate}
    \item\label{item:loground_phase-1-step-3} Consider a cluster $c\in \cC^{(i-1)} \setminus \cR^{(i)}$ that has \emph{no} neighbor in $\cR^{(i)}$.
        Let $C' \subseteq \cC^{(i-1)}$ be all the clusters of $\cC^{(i-1)}$ in the \emph{neighborhood} of $c$.
        For each $c_j \in C'$ move the lowest weight edge in $\rE(c, c_j)$ to $E_S$ and discard all edges in $\rE(c, c_j)$ from $\rE$.
    \item\label{item:loground_expanding_clusters} The clustering $\cC^{(i)}$ is formed by taking the clusters in $\cR^{(i)}$,
        and then extending them using all the edges in $\cE^{(i)}$
        to absorb other clusters that are connected to $\cR^{(i)}$ (using only edges in $\cE^{(i)}$).

        \begin{itemize}
            \item Specifically, let $c\in \cR^{(i)}$ be a sampled cluster, and let $\Delta^{(i-1)}(c)
                = \{\bar c\in \cC^{(i-1)}\mid E(c, \bar c)\cap\cE^{(i)}\not= \emptyset\}$ be the set of adjacent clusters that will be absorbed.
                Each such $c$ results a new cluster $c'\in \cC^{(i)}$, where $c'$ has the same root node as $c$,
                and the tree $T(c')$ is formed by attaching the trees $T(\bar c)$ (for each $\bar c\in \Delta^{(i-1)}(c)$),
                to the corresponding leaf node of the tree $T(c)$, using the appropriate edge in $E(c,\bar c)\cap \cE^{(i)}$
                (by construction, there is exactly one such edge).
        \end{itemize}
    \item\label{item:loground_phase-1-step-5} Remove all edges $(u,v)\in \rE$ where $u$ and $v$ belong to the same cluster in $\cC^{(i)}$.
        This set $\rE$ at the end of the $i^{th}$ epoch is denoted $\rE^{(i)}$.
        We then contract and form the new quotient graph (super-graph) and proceed to the next epoch.
\end{enumerate}

\subparagraph{Phase 2:} In the second phase, let $V'$ be the set of all endpoints of the remaining edges $\rE^{(\log k)}$.
For each $v\in V'$ and each $c\in \cC^{(\log k)}$, we add the lowest edge in $\rE(v, c)$ to $E_S$ before discarding the edges in $\rE(v, c)$.

\subsection{Analysis of Phase 1}
We first show that, for each edge $e = (u,v)$ that is discarded (not added to $E_S$),
there exists a path from $u$ to $v$ in $E_S$, of weight at most $k \cdot w_e$
(see \cref{thm:loground_removed_edges}), i.e., the edge $e$ is \emph{spanned} by existing spanner edges in $E_S$.
Next, in \cref{sec:loground_size_analysis}, we show that the number of edges added to the spanner $E_S$ during phase 1,
is $O(n^{1+1/k}\log k)$ in expectation (see \cref{thm:loground_number_of_added_edges}).

\subsubsection{Stretch Analysis}\label{sec:log_round_stretch}
We begin by providing some definitions used throughout the analysis, and then get into a formal stretch analysis.
\begin{definition}[Weighted-Stretch Radius]
\label{def:clustering_strong_radius}
$\cC$ is a clustering of weighted-stretch radius $r$ with respect to an edge set $\rE$ in a graph $G$ if and only if
\begin{enumerate}[(A)]
    \item\label{item:strong-radius-A} For all $c\in \cC$, the cluster $c$ has radius at most $r$ (equivalently, $T_c$ has depth at most $r$).
    \item\label{item:strong-radius-B} For each edge $e=(x,v)\in \rE$ such that $x\in c\in \cC$,
    all edges on the path from $x$ to the root of $T_c$ have weight less than or equal to $w_e$.
\end{enumerate}
\end{definition}
\begin{definition}[Cluster of a vertex]
\label{def:loground_cluster_vertex}
For a vertex $v$, $\cc^{(i)}(v)$ refers to the cluster of $\cC^{(i)}$ containing $v$.
\end{definition}

\begin{definition}[Cluster center]
\label{def:loground_cluster_center}
For a vertex $v$, $\cF^{(i)}(v)$ denotes the center of $\cc^{(i)}(v)$.
\end{definition}

First, an inductive argument shows that all the remaining edges in $\rE^{(i)}$ are between the current set of clusters $\cC^{(i)}$. 

\begin{restatable}{lemma}{logroundRemainingEdges}
\label{lem:loground_remaining_edges}
During the execution of Phase 1, any edge $e = (u,v)\in \rE$ at the end of epoch $i$ is such that
both end-points are members of distinct clusters in $\cC^{(i)}$.
\end{restatable}

\begin{proof}
We prove this statement by induction.

\textbf{Base case:} Before the first epoch, $\rE = E$ and all the edges have endpoints in $\cC_0$ since this is the set of all vertices.

\textbf{Inductive hypothesis:}
Assume that at the beginning of epoch $i$, all edges in $\rE$ have both endpoints in distinct clusters of $\cC^{(i-1)}$. 

\textbf{Induction:}
Towards a contradiction, assume that there is an edge $e=(u,v)\in \rE$ that survives to the end of epoch $i$ and has at least one endpoint that is not in any cluster of $\cC^{(i)}$. Without loss of generality, assume that this endpoint is $v$. Note that $\cc^{(i-1)}(v)$ and $\cc^{(i-1)}(u)$ exists by the inductive hypothesis. (See \cref{def:loground_cluster_vertex} for the definition of $\cc^{(i-1)}(\cdot)$.)

If $\cc^{(i-1)}(v)$ is adjacent to any cluster in $\cR^{(i)}$, then it would have been processed in \cref{item:loground_phase-1-step-2}.
Therefore, some edge between $\cc^{(i-1)}(v)$ and $r \in \cR^{(i)}$ was added to $\cE^{(i)}$.
In this case, $\cc^{(i-1)}(v)$ will be absorbed into a new cluster in $\cC^{(i)}$ (see \cref{item:loground_expanding_clusters}), and consequently, $v$ is also a member of $\cC^{(i)}$. Hence, $\cc^{(i-1)}(v)$ was not adjacent to any cluster in $\cR^{(i)}$.
\\
So, $\cc^{(i-1)}(v)$ was processed in \cref{item:loground_phase-1-step-3}.
In this case, all edges in $\rE(\cc^{(i-1)}(v), \cc^{(i-1)}(u))$ were discarded (one of the edges was added to the spanner $E_S$), and hence this case could not happen neither. This now leads to a contradiction (as \cref{item:loground_phase-1-step-2} or \cref{item:loground_phase-1-step-3} has to occur) and implies that both $u$ and $v$ belong to some clusters in $\cC^{(i)}$.

It remains to show that $u$ and $v$ belong to distinct clusters of $\cC^{(i)}$. But this follows directly from \cref{item:loground_phase-1-step-5}.
\end{proof}
Next, we argue inductively that in each epoch the cluster radius grows by a factor of $3$. 
\begin{restatable}{theorem}{logroundClusterRadius}
\label{thm:loground_cluster_radius}
At the end of epoch $i$, $\cC^{(i)}$ is a clustering of weighted-stretch radius (see Definition \ref{def:clustering_strong_radius})
at most $\frac{3^i-1}{2}$ with respect to the current set $\rE^{(i)}$.
\end{restatable}

\begin{proof}
We first prove Property~\eqref{item:strong-radius-A} and then Property~\eqref{item:strong-radius-B} of \cref{def:clustering_strong_radius}.
Each of the properties are proved by an inductive argument.

\subparagraph{Property~\eqref{item:strong-radius-A} of Definition \ref{def:clustering_strong_radius}.}
As a base case, notice that before the first epoch, each cluster has radius $0$.

During epoch $i$, each cluster $c'\in\cC^{(i)}$ is a union of a cluster $c\in \cR^{(i)}\subseteq \cC^{(i-1)}$,
and some number of clusters in $\cC^{(i-1)}$ that are adjacent to $c$.
The root of cluster $c'$ remains the same as the root of $c$.
However, the radius of $c'$ becomes the depth of the new rooted tree,
which can be at most three times the old radius plus one (for the edge connecting the adjacent cluster).
Thus the radius (not necessarily strong) of the new larger cluster is at most $3r+1$, where $r$ is the radius of $\cC^{(i-1)}$.
For an illustration, see \cref{fig:loground_strong_radius_induction} and consider the distance between the center of cluster $c$ and vertex $x$.
Assuming that the inductive hypothesis is satisfied for $\cC^{(i-1)}$ i.e., $r\le \frac{3^{i-1}-1}{2}$,
we see that the radius of $\cC^{(i)}$ is at most $\frac{3^i-1}{2}$.

Note that this proof implies that Property~\eqref{item:strong-radius-A} holds independently of Property~\eqref{item:strong-radius-B}.

\subparagraph{Property~\eqref{item:strong-radius-B} of Definition \ref{def:clustering_strong_radius}.}
As an inductive hypothesis, assume that the clustering $\cC^{(i-1)}$ has weighted-stretch radius $\frac{3^{i-1}-1}{2}$ with respect to $\rE^{(i-1)}$.
Now, consider an edge $e=(x,y)\in \rE^{(i)}$, such that $x\in c'\in \cC^{(i)}$ and $y\not\in c'$.
Note that by \cref{lem:loground_remaining_edges}, each edge $e \in \rE^{(i)}$ is of this form,
i.e., the endpoints of $e$ belong to distinct clusters of $\cC^{(i)}$.
According to \cref{item:loground_expanding_clusters}, cluster $c'$ was formed from a sampled cluster $c\in \cR^{(i)}$
that \emph{engulfed} the adjacent clusters in $\Delta^{(i-1)}(c)$
\begin{figure}[htbp]
    \centering
    \includegraphics[width=0.75\textwidth]{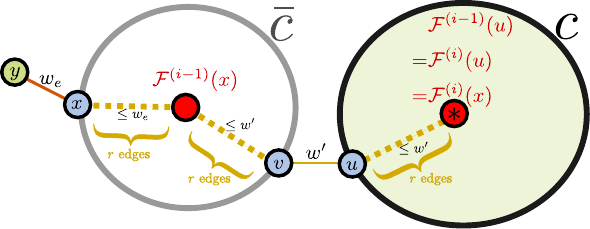}
    \caption{Sampled super-node $c$ will engulf $\bar c$, which is also a supernode (not sampled).
    Both correspond to clusters of radius $r$.}
    \label{fig:loground_strong_radius_induction}
\end{figure}

\subparagraph{Case $x\in \bar c\in \Delta^{(i-1)}(c)$:}
Let $w'$ be the weight of the edge $(u,v)\in \cE^{(i)}$ that connects $u\in T_c$ to $v\in T_{\bar c}$,
and let $w_e$ be the weight of the edge $e=(x, y)$ (see \cref{fig:loground_strong_radius_induction}).
We construct a path from $x$ to $\cF^{(i)}(x)$, by concatenating the following paths
$x\textcolor[rgb]{0.83,0.67,0.00}\patharrow \cF^{(i-1)}(x)\textcolor[rgb]{0.83,0.67,0.00}
\patharrow v\textcolor[rgb]{0.83,0.67,0.00}\patharrow u\textcolor[rgb]{0.83,0.67,0.00}\patharrow \cF^{(i)}(u) = \cF^{(i)}(x)$.
Since we know that the edge $e=(x,y)$ survived \cref{item:loground_phase-1-step-2-step-ii}, it must have weight at least as the edge $(u,v)$, i.e., $w' \le w_e$.
By the inductive hypothesis for level $i-1$, all edges on the first two segments of the above path have weight at most $w'$.
Similarly, the first two segments of the path only contain edges with weight at most $w_e$ (again using the inductive hypothesis at $i-1$).
Furthermore, the number of edges on the path is at most $r+r+1+r = 3r+1$, where $r$ is the weighted-stretch radius of clustering at the previous epoch.

\subparagraph{Case $x\in c$:}
Since $e=(x,y)\in \rE^{(i)}\implies (x,y)\in \rE^{(i-1)}$,
the inductive hypothesis implies that the path from $x$ to the root of $T_c$ (the center is $\cF^{(i-1)}(x) = \cF^{(i)}(x)$)
only uses edges of weight at most $w_e$.
\end{proof}

\begin{restatable}{corollary}{logroundFinalClusterRadius}
\label{cor:loground_final_cluster_radius}
The weighted-stretch radius of the final clustering $\cC^{(\log k)}$ is $\frac{k^{\log 3}-1}{2}$.
\end{restatable}

Using Theorem \ref{thm:loground_cluster_radius}, we show the following. 
\begin{restatable}{theorem}{logroundRemovedEdges}
\label{thm:loground_removed_edges}
For all edges $e = (u,v)$ removed in Phase 1, there exists a path between $u$ and $v$ in $E_S$ of weight at most $w_e\cdot k^{\log 3}$.
\end{restatable}

\begin{proof}
Depending on when $e = (u,v)$ was removed from $\rE$, there are three cases to consider: \cref{item:loground_phase-1-step-2}, \cref{item:loground_phase-1-step-3} and \cref{item:loground_phase-1-step-5}. Before analyzing these cases, note that by \cref{lem:loground_remaining_edges} after a cluster is formed no edge is removed from it in the subsequent epochs. This fact will be important in the rest of this proof as when we show that there is a path between $u$ and $v$ of weight at most $k^{\log{3}} \cdot w_e$, we will show that this path belongs to a cluster (or to two adjacent ones). Hence, once this path belongs to a cluster it also belongs to $E_S$.  \subparagraph{Case $e$ was removed in \cref{item:loground_phase-1-step-2} in epoch $i$.} Let $e_1 = (x, y)$ be the edge that was kept between $\cc^{(i-1)}(u)$ and $\cc^{(i-1)}(v)$. Let $x \in \cc^{(i-1)}(u)$ and $y \in \cc^{(i-1)}(v)$. Then, by \cref{thm:loground_cluster_radius} there exists a path from $u$ to $x$ within $\cc^{(i-1)}(u)$ of weight at most $2 \tfrac{3^{i-1} - 1}{2} \cdot w_e$. Similarly, there is a path between $v$ and $y$ within $\cc^{(i-1)}(v)$ of weight at most $2 \tfrac{3^{i-1} - 1}{2} \cdot w_e$. Since $E_S$ also contains the edge $e_1$ and $w_{e_1} \le w_e$ by \cref{item:loground_phase-1-step-2-step-i,item:loground_phase-1-step-2-step-ii}, we have that $E_S$ contains a path between $u$ and $v$ of weight at most $(2 (3^{i-1} - 1) + 1) \cdot w_e < 3^i \cdot w_e \le k^{\log{3}} w_e$.  \subparagraph{Case $e$ was removed in \cref{item:loground_phase-1-step-3} in epoch $i$.} This case is analogous to the previous one.  \subparagraph{Case $e$ was removed in \cref{item:loground_phase-1-step-5} in epoch $i$.} Let $c \in \cC^{(i)}$ be the cluster from which $e = (v_1, v_2)$ was removed. By construction, this edge is between two clusters $c_1$ and $c_2$ from $\cC^{(i - 1)}$ that are merged with $c$ in this epoch. Assume that $v_1 \in c_1$. Let $e_1 = (x_1, x)$ be the edge via which $c_1$ was merged to $c$, and let $x_1 \in c_1$. Also, let $e_2 = (y_2, y)$ be the edge with which $c_2$ was merged with $c$, and let $y_2 \in c_2$.  Since $e$ was not removed before \cref{item:loground_phase-1-step-5}, it means that when $c_1$ and $c_2$ got merged with $c$ the edge $e$ was not discarded in \cref{item:loground_phase-1-step-2-step-ii}. This in turn implies that $w_{e} \ge w_{e_1}$ and $w_e \ge w_{e_2}$.

Now, similar to the analysis of \cref{item:loground_phase-1-step-2}, using \cref{thm:loground_cluster_radius} we have that there is a path in $E_S$ between $v_1$ and $x_1$ of weight at most $(3^{i - 1} - 1) \cdot w_e$. Also, there is a path between $x$ and $y$ in $E_S$ of weight at most $(3^{i - 1} - 1) \cdot w_{e_1} \le (3^{i - 1} - 1) \cdot w_{e}$. Finally, there is a path between $v_2$ and $y_2$ in $E_S$ of weight at most $(3^{i - 1} - 1) \cdot w_e$. Combining these together, $E_S$ contains a path between $v_1$ and $v_2$ of weight at most $(3 \cdot (3^{i - 1} - 1) + 2) \cdot w_e = (3^i - 1) \cdot w_e \le k^{\log{3}} \cdot w_e$, as desired.
\end{proof}

\subparagraph{Stretch Analysis of Phase~2.}

Recall, that in the second phase, we let $V'$ be the set of all endpoints of the \emph{un-processed} edges in $\rE^{(\log k)}$.
Subsequently, we add the lowest edge in $\rE(v, c)$ to $E_S$ before discarding the edges in $\rE(v, c)$, for each $v\in V'$ and $c\in \cC^{(\log k)}$.

Using the weighted stretch radius of the final clustering from \cref{cor:loground_final_cluster_radius},
we can prove the following lemma, using an argument similar to \cref{thm:loground_removed_edges}.
We omit the formal proof to avoid repetition.
\begin{restatable}{lemma}{logroundPhase_2RemovedEdges}
\label{lem:loground_phase_2_removed_edges}
For each edge $w_e = (v, c)\in V'\times \cC^{(\log k)}$ removed in Phase~2,
there exists a path between $u$ and $v$ in $E_S$ of weight at most $w_e\cdot k^{\log 3}$.
\end{restatable}

\subsubsection{Size Analysis}
\label{sec:loground_size_analysis}
Next, we provide an upper-bound on $E_S$.
First, we upper-bound the number of clusters in each $\cC^{(i)}$.
\begin{restatable}{lemma}{logroundNumberOfClusters}
\label{lem:loground_number_of_clusters}
For each $i \le\log{k}$, in expectation it holds $|\cC^{(i-1)}| \in O\rb{n^{1 -\frac{2^{i-1} - 1}{k}}}$.
\end{restatable}
\begin{proof}
A cluster $c$ belongs to $\cC^{(i-1)}$ only if $c$ was sampled to $\cR_j$ in \cref{item:loground_phase-1-step-1} for each $1 \le j \le i-1$.

 This happens with probability $\prod_{j = 1}^{i-1} n^{-\frac{2^{j - 1}}{k}} = n^{-\frac{2^{i-1} - 1}{k}}$.
 Therefore, $\ee{|\cC^{(i-1)}|} = n^{1 -\frac{2^{i-1} - 1}{k}}$.
\end{proof}

Building on \cref{lem:loground_number_of_clusters} we obtain the following claim.
\begin{restatable}{theorem}{logroundNumberOfAddedEdges}
\label{thm:loground_number_of_added_edges}
During Phase~1, in expectation there are $O\rb{n^{1+1/k} \cdot \log k}$ edges added to $E_S$.
\end{restatable}
\begin{proof}
    \cref{item:loground_phase-1-step-1,item:loground_expanding_clusters,item:loground_phase-1-step-5} do not affect $E_S$. Hence, we analyze only the remaining steps.
	

		Consider epoch $i$. Fix a cluster $c\in \cC^{(i-1)}$ (which might or might not be in $\cR^{(i)}$). We will upper-bound the number of edges that in expectation are added when considering $c$.

		Let $p = n^{-\frac{2^{i-1}}{k}}$. Recall that each cluster $c'$ is added from $\cC^{(i-1)}$ to $\cR^{(i)}$ independently and with probability $p$.
		Order the clusters $c'$ of $\cC^{(i-1)}$ adjacent to $c$ in the non-decreasing order by the lowest-edge in $\rE(c, c')$.
		Consider the first $A$ among those sorted clusters. 
		
		Taking into account both \cref{item:loground_phase-1-step-2,item:loground_phase-1-step-3}, an edge from $c$ to the $A$-th cluster is added to $E_S$ if and only if all previous clusters are not sampled, which happens with probability $(1-p)^{A-1}$. Hence, the expected number of edges added by $c$ is upper-bounded by $\sum_{A=1}^{n} (1-p)^{A-1} <\sum_{A=1}^{\infty} (1-p)^{A-1} = \frac{1}{p}.$
		Hence, we have that in expectation the number of edges added to the spanner when considering $c$ is $O(1/p).$

\end{proof}

\subparagraph{Size Analysis of Phase 2.} Finally, As a corollary of \cref{lem:loground_number_of_clusters}, we see that in expectation $|\cC^{(\log k)}| \in O\left(n^{1/k}\right)$.
Therefore, the number of edges added in Phase~2 is at most $|V'|\cdot|\cC^{(\log k)}| \in O\left(n^{1+1/k}\right)$. This concludes our spanner construction, and yields the main theorem.

\begin{theorem}
Given a weighted graph $G$, the cluster-merging algorithm builds a spanner of stretch $O(k^{\log{3}})$ and expected size $O(n^{1+1/k} \cdot \log{k})$, within $O(\log{k})$ epochs.
\end{theorem}


\section{General Trade-off between Stretch and Number of Rounds}
\label{sec:general}

In this section, we provide an overview of an algorithm that combines ideas of \cref{sec:polyround_algorithm} and \cref{sec:loground_cluster_merging}, with the cluster-vertex merging concept that we described earlier.
This gives us a general tradeoff between number of rounds, and stretch.
For instance, we can construct a spanner with stretch $k^{1+o(1)}$ in $\frac{\log^2(k)}{\log \log(k)}$ rounds.
At a high-level, the algorithm runs in a sequence of epochs, and each epoch performs $t$ iterations of \cite{baswana_sen}.

We can imagine the algorithm of \cite{baswana_sen} as being one extreme of this tradeoff (when $t=k$).
The algorithm of \cref{sec:polyround_algorithm} generalizes this, by splitting the $k$ iterations of \cite{baswana_sen} over two epochs,
each with $\sqrt k$ iterations.
After the first set of $\sqrt{k}$ iterations (the first epoch), we \emph{contract the most recent clusters},
and then repeat $\sqrt{k}$ iterations (the second epoch) on the contracted graph.
Importantly, the second epoch uses different sampling probabilities --
as if we were actually trying to construct a stronger $\mathcal O(\sqrt k)$ stretch spanner.

Meanwhile, the algorithm of \cref{sec:loground_cluster_merging} occupies the other extreme of our tradeoff.
this algorithm immediately contracts after a single ``\cite{baswana_sen}-like step''.
Consequently, each step now becomes an ``epoch'', and they all use different sampling probabilities.
With $\log k$ epochs, this is the \emph{``fastest''} algorithm in our tradeoff, and thus has the worst stretch.

We interpolate between these extremes by repeating the following two steps:
\begin{itemize}
    \item In each epoch, we grow clusters, based on the cluster-vertex merging approach, till a certain radius $t$ on the \textit{quotient} graph,
    where each \emph{super-node} in the graph is a contracted cluster from the previous epoch.
    \item At the end of an epoch, we contract the clusters of radius $t$ to obtain the quotient graph for the next epoch,
    adjust the sampling probabilities and continue.
\end{itemize}
The parameter $t$ can be varied, thus resulting in a family of algorithms that achieve our trade-offs.
For instance, $t=1$ corresponds to \cref{sec:loground_cluster_merging}, $t=\sqrt k$ corresponds to \cref{sec:polyround_algorithm},
and $t=k$ brings us back to the algorithm of \cite{baswana_sen}.

The idea behind the generalization is that now, the stretch (equivalently, the radius of clusters),
rather than increasing by a \textit{multiplicative} factor of $3$ in each epoch (as it did in \cref{sec:loground_cluster_merging}),
now it grows by a factor of $(2t+1)$ in each epoch.
We again use the intuition that after each contraction, since the remaining graph is smaller in size,
we can afford to grow clusters faster, and we adjust this rate by decreasing the sampling probabilities.


\subsection{General Trade-off Algorithm}

Let us first formalize the concept of a quotient graph before getting into the algorithm details:
\begin{definition}
\label{def:quotient_graph}
Given a graph $G=(V,E)$ and a clustering $\cC$, the quotient graph $\hat{G}=(\cC,\hat{E})$, is defined as follows.\footnote{The quotient graph is sometimes called a super-graph.} Its vertices are the clusters, and there is an edge between two clusters $c, c'\in \cC$, if there exist $v' \in c, v' \in c'$ such that $(v,v') \in E$.
\end{definition}
We next give a formal description of the algorithm. As can be expected, many of the steps inside each epoch is similar to the algorithms of previous sections and \cite{baswana_sen}.
\label{sec:general_algorithm}
\subparagraph{Phase 1:}
The first phase of the algorithm proceeds in $l = \log k/\log(t+1)$ \emph{epochs}.
Through the epochs, we create a sequence of clusterings $\cD^{(1)}, \cD^{(2)},\cdots, \cD^{(l)}$,
and graphs $G^{(0)}\supset G^{(1)}\supset\cdots\supset G^{(l)}$.
Here $G^{(0)} = G$, and $G^{(i)} = (V^{(i)}, E^{(i)}) = \nicefrac{G^{(i-1)}}{\cD^{(i)}}$
is the quotient graph induced by the $i^{th}$ level clustering.

In what follows, we will use the term \emph{vertex} to refer to a member of $V = V^{(0)}$ i.e. a vertex of the original graph.
On the other hand, a \emph{super-node} will refer to a member of any intermediate $V^{(i)}$ (including $V^{(0)}$).

Throughout the execution, we maintain a set of edges $\rE$.
The state of this set at the end of epoch $i$ is denoted as $\rE^{(i)}$, and $\rE^{(0)}$ is initialized to all the edges in $G$.
During epoch $i$, edges are removed from $\rE^{(i-1)}$ to form $\rE^{(i)}$.
Some of these removed edges are added to the set of spanner edges $E_S$, and the other ones are discarded.

At epoch $i$ we perform the following steps:
\begin{enumerate}[(A)]
    \item\label{item:general-step-BS} Let $G^{(i-1)} = (V^{(i-1)}, E^{(i-1)})$ be the graph at this epoch,
        and let $\cD^{(i)}_0$ be the initial clustering comprising of all singleton \emph{super-nodes}
        i.e. $\cD^{(i)}_0 \eqdef \{ \{ v\}\mid v\in G^{(i)}\}$.

        Next (in \cref{item:general-step-contraction}), we perform $t$ iterations of Baswana-Sen\footnote{with fixed sampling probability $n^{-\frac{(t+1)^{i-1}}{k}}$} on $G^{(i)}$,
        in order to obtain the final clustering $\cD^{(i)}_t$.
        We also maintain a set of edges $\cE^{(i)}_j \subseteq E_S$ at iteration $j$,
        where the edges in $\cE^{(i)}_j$ induce the clustering at the $j^{th}$ iteration.
        Henceforth, we will drop the superscript $(i)$ when it is clear from context.

        We will use the set $\rE^{(i)}_j$, (where $\rE^{(i)}_0$ is initialized to $\rE^{(i-1)}$),
        to denote the set of unprocessed edges at the end of iteration $j$.
        When clear from context, we will abbreviate this as $\rE$.
        \label{item:initialize_epoch}
    \item\label{item:general-step-contraction} For $j = 1,2,\cdots,t$, iteration $j$ proceeds as follows:
    \begin{enumerate}[1.]
        \item Sample a set of clusters $\cR_j \subseteq \cD_{j-1}$,
            where each $c\in \cD_{j-1}$ is chosen to be a member of $\cR_j$ with probability $n^{-\frac{(t+1)^{i-1}}{k}}$.
            We also initialize $\cE_j$ to the subset of edges in $\cE_{j-1}$ that are contained in some cluster $c\in \cR_j$.
            \label{item:subsample_clusters}
        \item Consider all (contracted) \emph{super-nodes} $v\in G^{(i-1)} \setminus \cR_j$ that are not in any sampled cluster.
            We define $\Gamma_j(v) \subseteq \cD_{j - 1}$ to be all the clusters of $\cD_{j - 1}$ in the \emph{neighborhood} of $v$.
        \item\label{item:general-step-B-has-neighbor} For each super-node $v$, such that $\Gamma_j(v)$ intersects $\cR_j$
            (i.e. super-nodes with some neighboring sampled cluster),
            let $\cN_j(v)\in \cR_j\cap \Gamma_j(v)$ be the \emph{closest neighboring sampled cluster} of $v$.
            \label{item:neighboring_sampled_cluster}
        \begin{itemize}\label{item:general-step-B-no-neighbor}
            \item Add the lowest weight edge $e$ in $\rE(v, \cN_j(v))$
                to both $\cE_j$ and $E_S$ and remove the entire set $\rE(v, \cN_j(v))$ from $\rE$.
            \item For all clusters $c\in \Gamma_{j}(v)$ adjacent to $v$ with any edge of weight strictly less than $e$,
                Add the lowest weight edge in $\rE(v, c)$ to $E_S$ and then discard all the edges in $\rE(v, c)$ from $\rE$.
        \end{itemize}
        \item For all super-nodes $v$, such that $\Gamma_j(v)\cap \cR_j = \emptyset$ (i.e. super-nodes with no neighboring sampled cluster),
            \label{item:no_neighboring_sampled_cluster}
            \begin{itemize}
                \item For each $c \in \Gamma_j(v)$ move the lowest weight edge in $\rE(v, c)$ to $E_S$ and discard all edges in $\rE(v, c)$ from $\rE$.
            \end{itemize}
        \item The clustering $\cD_j$ is formed by taking each cluster in $\cR_j$, and then extending it using the edges in $\cE_j$,
            to absorb other \emph{super-nodes} that are connected to $\cR_j$ (using only edges in $\cE_j$).

            Specifically, for each $c\in \cR_j$, we define $\Delta_j(c) = \{v\in V^{(i-1)}\mid E^{(i-1)}(v, c)\cap\cE_j\not= \phi\}$
            as the set of adjacent \emph{super-nodes} that will be absorbed.
            Each such $c$ results a new cluster $c'\in \cD_j$, where $c'$ has the same root super-node as $c$,
            and the tree $T(c')$ is formed by attaching each \emph{super-node} $v\in \Delta_{j-1}$ to the corresponding leaf node of $T(c)$,
            using the appropriate edge in $E(c,\bar c)\cap \cE_j$ (by construction, there is exactly one such edge).

            Note that this description is simpler than the one presented in \cref{item:loground_expanding_clusters} in \cref{sec:loground_algorithm},
            because we are working in the intermediate graph $G^{(i)}$.
        \item Remove all edges $(u,v)\in \rE$ where $u$ and $v$ belong to the same cluster in $\cD_j$.
            This set $\rE$ at the end of the $j^{th}$ iteration (in the $i^{th}$ epoch) is denoted by $\rE^{(i)}_j$.
            \label{item:remove_intercluster_edges}
    \end{enumerate}
    \item Construct the quotient graph $G^{(i+1)} = \nicefrac{G^{(i)}}{\cD^{(i)}_t}$.
        For all pairs of \emph{super-nodes} $u,v\in V^{(i)}$,
        we only keep the minimum weight edge between $u$ and $v$ in $\rE$ (currently denoted by $\rE^{(i)}_t$).
        All other edges are discarded, and the final set at the end of the epoch is denoted by $\rE^{(i)}$.
        \label{item:contract_clusters}
\end{enumerate}

\subparagraph{Phase 2:}
In the second phase, let $V'$ be the set of all endpoints of the remaining edges $\rE^{(l)}$.
For each $v\in V'$ and $c\in \cD^{(l)}_t$, we add the lowest edge in $\rE(v, c)$ to $E_S$ before discarding the edges in $\rE^{(l)}(v, c)$.

\remove{For lack of space, the full details of the analysis appear in Section \ref{sec:omitted_proofs_from_sec_general}.}


\subsection{Intermediate Graph Clusterings on Original Graph $G$}
\label{sec:compose_clusterings}
Notice that the clusterings $\cD^{(i)}_j$ are defined on the intermediate graphs $G^{(i)}$.
For the sake of clarity, we define the clustering $\cC^{(i)}_j$, which is just the same $\cD^{(i)}_j$, but on the original graph $G = G^{(0)}$.
We achieve this by defining an operation that is essentially the inverse of the contraction being performed in \cref{item:contract_clusters}.
This operation allows us to define $\cC^{(i)}_j$ recursively as follows: 
\begin{definition}[Composition of Clusterings]
\label{def:clustering_composition}
Given a clustering $\cC^{(i-1)}$ defined on $G^{(0)}$,
and clustering $\cD^{(i)}_{j}$ defined on the contracted graph $G^{(i)} = \nicefrac{G}{\cC^{(i)}}$, we can obtain the corresponding $\cC^{(i)}_{j}$,
by replacing each super-node $v\in \cc_v\in \cD^{(i)}_j$ by the corresponding (older) cluster $\cc^{(i-1)}_v\in \cC^{(i-1)}$
(i.e. the super-node $v\in G^{(i)}$ was formed by contracting a cluster $\cc^{(i-1)}_v\in \cC^{(i-1)}$ in the original $G$).
Specifically, we modify the tree $T(\cc_v)$, by replacing each internal \emph{super-node} $v$ by the corresponding tree $T\left(\cc^{(i-1)}_v\right)$.
Thus, the final tree (also the cluster) only contains vertices from $G = G^{(0)}$.
\end{definition}

We also need to specify how we obtain $\cC^{(i)}$, and the following observation suffices.
\begin{definition}
\label{def:clustering_base_case}
Since $\cC^{(i)}_t$ corresponds to $\cD^{(i)}_t$, and $G^{(i+1)} = \nicefrac{G^{(i)}}{\cD^{(i)}_t}$,
and therefore, we define $\cC^{(i)}$ to be the same as $\cC^{(i)}_t$.
As a base case, $\cC^{(0)} \eqdef \cD^{(0)}$ (singleton clusters).
\end{definition}

In the next section, we will see how the radius of the clusters is affected by the composition defined above.
This should be reminiscent of \cref{thm:loground_cluster_radius}.
First, we start with some definitions towards proving an analog of \cref{lem:loground_remaining_edges}.

\begin{definition}[Cluster of a vertex]
\label{def:cluster_vertex}
For a vertex $v$ in the original vertex set $V$, $\cc^{(i)}_j(v)$ refers to the cluster of $\cC^{(i)}_j$ containing $v$.
\end{definition}

\begin{definition}[Cluster center]
\label{def:cluster_center}
For a vertex $v$, $\cF^{(i)}_j(v)$ denotes the center of $\cc^{(i)}_j(v)$.
\end{definition}

\begin{restatable}{lemma}{remainingEdges}
\label{lem:remaining_edges}
During the execution of Phase 1, for any un-processed edge $e = (u,v)\in \rE^{(i)}_j$ at the end of iteration $j$ in epoch $i$,
both end-points $u, v$ are members of distinct clusters in $\cC^{(i)}_j$.
\end{restatable}
\begin{proof}
The proof is essentially a combination of \cref{lem:loground_remaining_edges} and \cref{lem:polyround_remaining_edges},
and we proceed by induction on $j$.
In the case that $i\not= 0, j = 0$, we note that the clusters in $\cC^{(i)}_0$ correspond exactly to the clusters in $\cC^{(i-1)}_t$.
The edge set $\rE^{(i)}_0$ is also just $\rE^{(i-1)}_t$ under a re-labeling, and thus induction on $j$ suffices.

\textbf{Base case:} Initially, $\rE^{(0)}_0 = E$, and all the edges have endpoints in $\cC^{(0)}_0$,
since it is comprised of singleton clusters $\{ v\}$ for all vertices $v\in V^{(0)} = V$ (see \cref{item:initialize_epoch}).

\textbf{Inductive hypothesis:}
Assume that at the beginning of iteration $j$ within epoch $i$,
all edges in $\rE^{(i)}_{j-1}$ have both endpoints in distinct clusters of $\cC^{(i)}_{j-1}$.

In what follows, we will be working on the graph $G^{(i)}$ during the $i^{th}$ epoch,
and in general, we will drop the $(i)$ super-script from all symbols.

\textbf{Induction:}
Towards a contradiction, assume that there is an edge $e=(u,v)\in \rE_j$ that survives to the end of iteration $j$,
and has at least one endpoint that is not in any cluster of $\cC_j$.
Without loss of generality, assume that $v$ is such a ``problematic'' endpoint.
Note that since $\rE_j\subseteq \rE_{j-1}\implies (u,v)\in \rE_{j-1}$,
the inductive hypothesis implies that both $\cc_{j-1}(v)$ and $\cc_{j-1}(u)$ exist (see \cref{def:cluster_vertex}).

If the \emph{super-node} containing the ``problematic'' $v$ is adjacent to any cluster in $\cR_j$,
then it would have been processed in \cref{item:neighboring_sampled_cluster}.
Therefore, some edge between $\cc_{j-1}(v)$ and $\cN_j(v) \in \cR_j$ would have been added to $\cE_j$.
In this case, the \emph{super-node} containing $v$ would be absorbed into a new cluster in $\cC_j$, i.e. $v\in c\in \cC_j$.
Hence, in order to maintain the assumption that $v$ is ``problematic'', it must be the case that $v$ was not adjacent to any cluster in $\cR_j$.

What if the super-node containing $v$ was processed in \cref{item:no_neighboring_sampled_cluster}?
In this case, all edges in $\rE(v, \cc_{j-1}(u))$ were discarded (one of the edges was added to the spanner $E_S$), and hence $(u, v)\not\in \rE_j$.

This implies that neither \cref{item:neighboring_sampled_cluster} nor \cref{item:no_neighboring_sampled_cluster} can have occurred,
and therefore $v$ must have been a member of some cluster in $\cR_j$, which in turn implies that $v$ must be a member of some $c\in \cC_j$.
This contradicts the assumption that $v$ was ``problematic''.

Thus, we can conclude that both $u$ and $v$ belong to some clusters in $\cC_j$.
It remains to show that $u$ and $v$ belong to \emph{distinct} clusters of $\cC_j$, which follows directly from \cref{item:remove_intercluster_edges}.
\end{proof}

\subsection{General Stretch Analysis}
In this section, we provide a stretch analysis for our general algorithm. For this we need few more notations.

\begin{definition}
\label{def:strong_radius_variable}
For any epoch $i\in [l]$ and any $j\in [t]$,
we define $r^{(i)}_j$ to be the weighted-stretch radius of the clustering $\cC^{(i)}_j$ with respect to the edge set $\rE^{(i)}_j$.
For the sake of convenience, we also define $r^{(i)} = r^{(i+1)}_0 = r^{(i)}_t$.
\end{definition}

\begin{restatable}{lemma}{radiusRecurrence}
\label{lem:radius_recurrence}
For any $i\in [l]$ and any $j\in [t]$, we have $r^{(i)}_j \le r^{(i)}_{j-1} + 2r^{(i-1)} + 1$.
\end{restatable}
\begin{proof}
The proof proceeds analogous to the induction in \cref{thm:loground_cluster_radius},
where we claimed that the new radius of clustering is upper bounded by $r+2r+1$ ($r$ being the radius of the previous clustering).
This was illustrated in \cref{fig:loground_strong_radius_induction}.
It is strongly recommended that one reads the proof of \cref{thm:loground_cluster_radius} before this one, since we will be refer to it.

The difference in the general case, is that the clusters present at a certain iteration within the $i^{th}$ epoch,
do not have a radius that only depends on $i$.
Instead, we now also have a dependence on the iteration number $j$.
\begin{figure}[htbp]
    \centering
    \includegraphics[width=0.85\textwidth]{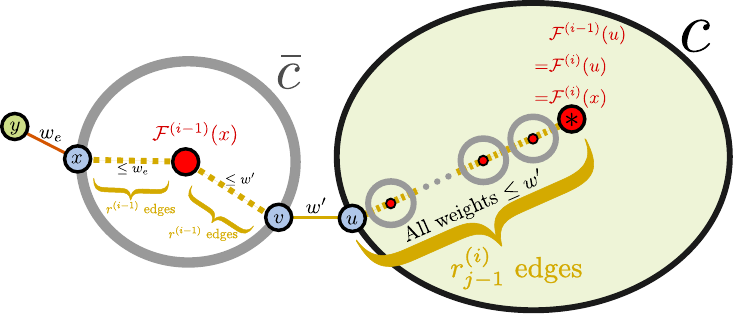}
    \caption{A figure used in the proof of \cref{lem:radius_recurrence}.
    Sampled super-node $c$ will engulf $\bar c$, which is also a supernode (not sampled).}
    \label{fig:strong_radius_induction}
\end{figure}

Specifically, following the proof of \cref{thm:loground_cluster_radius},
we consider a cluster $c$ of radius $r^{(i)}_{j-1}$ that is about to engulf a \emph{super-node} $\bar c$.
Recall that each such \emph{super-node} refers to a cluster in $\cD^{(i-1)}_t$ (or rather in $\cC^{(i-1)}_t$, since we are working in the original graph),
that was \emph{contracted} at the end of the $(i-1)^{th}$ epoch (\cref{fig:strong_radius_induction}), and thus has radius $r^{(i-1)}$.
The grey circles in \cref{fig:strong_radius_induction} correspond to such \emph{super-nodes},
each of which have weighted-stretch radius $r^{(i-1)}_t = r^{(i-1)}$ (this follows from inductive hypothesis).

To justify that $r^{(i)}_j$ satisfies the above recurrence,
we have to show that the two properties in \cref{def:clustering_strong_radius} hold for the clustering $\cC^{(i)}_j$.
The fact that Property~\eqref{item:strong-radius-A} holds, should be clear from \cref{fig:strong_radius_induction}.
Specifically, the radius of the cluster $\cC^{(i)}_j$ is at most one greater than the the sum of the radius of $\cC^{(i)}_{j-1}$
and \emph{twice} the radius of $\cC^{(i-1)}$.

In order to show that Property~\eqref{item:strong-radius-B} holds,
we consider an edge $e=(x,y)\in \rE^{(i)}_j$ such that $x\in c'\in \cC^{(i)}$, and $y\not\in c'$.
We can use an argument similar to the one in \cref{thm:loground_cluster_radius}
(\cref{fig:strong_radius_induction} illustrates the analog of the first case in the proof of \cref{thm:loground_cluster_radius}).
We omit the details to avoid repetition.
\end{proof}

\begin{restatable}{corollary}{epochClusterRadius}
\label{cor:epoch_cluster_radius}
For any $i\in [l]$, we have $r^{(i)} \le \frac{(2t+1)^i-1}{2}$.
\end{restatable}
\begin{proof}
We solve the recurrence from \cref{lem:radius_recurrence} by noticing the base case $r^{(0)} = 0$, and proceeding with induction.
\begin{align*}
    r^{(i)} = r^{(i)}_t &\le r^{(i)}_{t-1} + 2r^{(i-1)} + 1 \\
    &\le r^{(i)}_{t-2} + \left( 2r^{(i-1)} + 1\right) \\
    \implies r^{(i)} &\le r^{(i)}_0 + t\left( 2r^{(i-1)} + 1\right) = r^{(i-1)} + t\left( 2r^{(i-1)} + 1\right) \\
    &= (2t+1)\cdot r^{(i-1)} + t \\
    &= (2t+1)\cdot \frac{(2t+1)^{i-1}-1}{2} + t = \frac{(2t+1)^i-1}{2}
\end{align*}
\end{proof}

\begin{restatable}{corollary}{finalClusterRadius}
\label{cor:final_cluster_radius}
The final clustering $\cC^{(l)}$ has weighted-stretch radius $\frac{k^s - 1}{2}$ w.r.t. $\rE^{(l)}$, where $s = \frac{\log(2t+1)}{\log(t+1)}$
\end{restatable}
\begin{proof}
Substituting the value of $l = \frac{\log k}{\log (t+1)}$,
we see that $r^{(l)} = (2t+1)^l = (2t+1)^{\frac{\log k}{\log (t+1)}} = 2^{\frac{\log k\log(2t+1)}{\log (t+1)}} = k^{\frac{\log(2t+1)}{\log (t+1)}}$.
\end{proof}

Since \cref{def:clustering_strong_radius} only requires an upper bound on the radius,
the above \cref{cor:final_cluster_radius} implies that any intermediate clustering $\cC^{(i)}_j$ also has weighted-stretch radius $\frac{k^s - 1}{2}$.

\begin{restatable}{theorem}{removedEdges}
\label{thm:removed_edges}
For any edges $e=\{u,v\}$ of weight $w_e$ removed in Phase 1, there exists a path $\cP_{e} \in E_S$ of total weight at most $2k^s\cdot w_e$,
between $u$ and $v$, where $s = \frac{\log (2t+1)}{\log (t+1)}$.
\end{restatable}
\begin{proof}
The proof proceeds similarly to that of \cref{thm:loground_removed_edges}.
Edges are discarded in \cref{item:neighboring_sampled_cluster}, \cref{item:no_neighboring_sampled_cluster},
\cref{item:remove_intercluster_edges}, or \cref{item:contract_clusters}.

\subparagraph{Case $e (v, u)$ was removed in \cref{item:neighboring_sampled_cluster}, \cref{item:no_neighboring_sampled_cluster},
or \cref{item:contract_clusters}.}
While these removals are slightly different, they all share a common structure, which is illustrated in \cref{fig:weighted_stretch_path}.
Specifically, when removing an edge $e = (v,u)\in E^{(0)}$, the algorithm considers $c_v, c_u\in \cC^{(i)}$ such that $v\in c_v, u\in c_u$,
and it adds edge $e' = (v', u')\in \rE(c_v, c_u)$ to the spanner $E_S$, before discarding all the edges in $\rE(c_v, c_u)$ (including $e$).
Since $e'$ is always chosen to be the lowest weight edge, we have $w_{e'}\le w_e$.
We also define $\cF_v, \cF_u$ to be the centers of the $c_v, c_u$ respectively.
Assuming that $c_v$ and $c_u$ have \emph{weighted-stretch radius} $r_1, r_2$ respectively (see \cref{fig:weighted_stretch_path}),
we know that there is a path $\cF_v\textcolor[rgb]{0.83,0.67,0.00}\patharrow v'\textcolor[rgb]{0.83,0.67,0.00}\patharrow u'
\textcolor[rgb]{0.83,0.67,0.00}\patharrow \cF_u$ of weight at most $w_{e'}(r_1+1+r_2)$ that use only edges in $E_S$.
Similarly, we know that there are paths $v\textcolor[rgb]{0.83,0.67,0.00}\patharrow \cF_v$ and
$\cF_u\textcolor[rgb]{0.83,0.67,0.00}\patharrow u$ of total weight at most $w_e(r_1+r_2)$, again using only spanner edges.
\cref{fig:weighted_stretch_path} illustrates this construction.
So, we can conclude that the discarded edge of weight $w_e$ can be replaced by a path of total weight at most $w_e(2r_1+2r_2+1)$.
Since \cref{cor:final_cluster_radius} states that $r_1, r_2\le \frac{k^s-1}{2}$, we conclude that the \emph{stretch} of edge $e$ is at most $2k^s$.
\begin{figure}[htbp]
    \centering
    \includegraphics[width=0.75\textwidth]{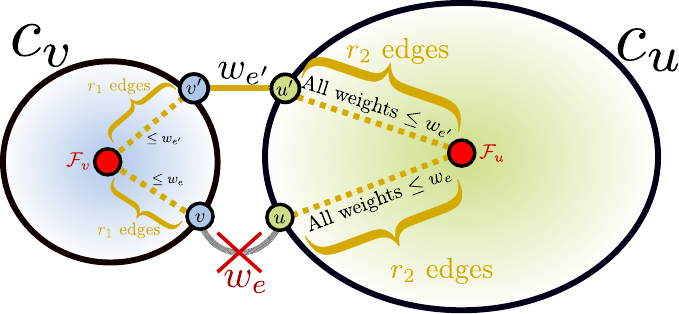}
    \caption{A figure used in the proof of \cref{thm:removed_edges}.}
    \label{fig:weighted_stretch_path}
\end{figure}

\subparagraph{Case $e=(v,u)$ was removed in \cref{item:remove_intercluster_edges} in epoch $i$, iteration $j$.}
Let $c \in \cC^{(i)}$ be the cluster from which $e$ was removed.
This means that both $v$ and $u$ were absorbed into the same sampled cluster $c_e\in \cR_j$ with center $\cF_e$.
Since $(v,u)$ was not added to the spanner at \cref{item:neighboring_sampled_cluster} while processing the \emph{super-node} containing $v$,
the edge $e' = (v, v')$ that was actually added must have weight $w_{e'}\le w_e$ (note that $v'\in c_e$).
Assuming that the cluster $c_e$ has \emph{weighted-stretch radius} $r$, we can conclude that there is a path $v\patharrow v'\patharrow \cF_e$,
of total weight $\le (1+r)\cdot w_e$, only using edges in $E_S$.
Similarly, since $(v,u)$ was also not added to $E_S$ while processing the super-node containing $u$,
we can conclude that there is a path $u\patharrow u'\patharrow \cF_e$, of total weight $\le (1+r)\cdot w_e$, only using edges in $E_S$.
Concatenating these two paths, we see that $e$ has stretch at most $(2r+2) \le k^s$.
\end{proof}

Finally, we can use a very similar argument to the stretch analysis in \cref{lem:loground_phase_2_removed_edges} for edges removed in Phase 2. It follows that the overall stretch of our spanner algorithm is $O(k^{s})$.
\subsection{General Size Analysis}
\label{sec:general_size_analysis}
We now upper-bound the size of a spanner constructed by the algorithm from this section.
The analysis presented here is in many similar to the one given in \cref{sec:loground_size_analysis}.
The following claim is similar to \cref{lem:loground_number_of_clusters}.
\begin{restatable}{lemma}{clusteringSize}
\label{lem:clustering_size}
At the end of epoch $i$, the number of super-nodes in the graph $G^{(i)}$ (which is the same as the number of clusters $\cD^{(i)}_t$),
is $O\rb{n^{1-\frac{(t+1)^i-1}{k}} + \log{n}}$ in expectation.
\end{restatable}
\begin{proof}
A cluster $c\in \cD^{(i-1)}_t$ belongs to $\cD^{(i)}_t$ only if $c$ was sampled to $\cR^{(i)}_j$
in \cref{item:subsample_clusters} for each $1\le j\le t$.
This happens with probability $\prod\limits_{j=1}^{t} n^{-\frac{(t+1)^{i-1}}{k}} = n^{-t\cdot \frac{(t+1)^{i-1}}{k}}$.
So, the probability that a specific \emph{vertex} $v\in V^{(0)}$ survives in $V^{(i)}$ is:
\[
\prob{v\in V^{(i)}} = \prod\limits_{i'=1}^{i} n^{-t\cdot \frac{(t+1)^{(i'-1)}}{k}} = n^{-\frac{(t+1)^{i} - 1}{k}}
\implies \mathbb E\left[ |V^{(i)}|\right] = n^{1 - \frac{(t+1)^{i} - 1}{k}}
\]
\end{proof}

\begin{restatable}{corollary}{finalClusteringSize}
\label{cor:final_clustering_size}
The expected number of super-nodes in the final contracted graph $G^{(l)}$ is $\mathcal O\left(n^{1/k}\right)$.
\end{restatable}

\begin{restatable}{lemma}{phaseOneNumAddedEdges}
\label{lem:phase_1_num_added_edges}
During Phase~1, in expectation there are $O\rb{n^{1 + 1/k} \cdot (l + t)}$ edges added to $E_S$.
\end{restatable}
\begin{proof}
    We first analyze the contribution of \cref{item:general-step-BS} and then the contribution of \cref{item:general-step-contraction} to the size of spanner across all the epochs.

    \subparagraph{Analysis of \cref{item:general-step-BS}.} 
    Let $n_i$ be the number of super-nodes at the beginning of epoch $i$. Then, by following \cite{baswana_sen} analysis, this step adds $O\rb{n_i^{1 + 1/k} \cdot t}$ edges to the spanner. So, by \cref{lem:clustering_size} we have that across all the epochs this step adds the following number of edges to the spanner
    \begin{align*}
        \sum_{i = 1}^l O\rb{n_i^{1 + 1/k} \cdot t} & = \sum_{i = 1}^l O\rb{n^{(1-\frac{(t+1)^{i - 1}-1}{k}) (1 + 1/k)} \cdot t} \\
        & \in O\rb{n^{1 + 1/k} \cdot t} \sum_{i = 1}^l n^{-\frac{(t+1)^{i - 1}-1}{k}} \\
        & = O\rb{n^{1 + 1/k} \cdot t}.
    \end{align*}

    \subparagraph{Analysis of \cref{item:general-step-contraction}.}
    Fix an epoch $i$. Observe that by letting $t = 1$, then this step corresponds to the algorithm given in \cref{sec:loground_algorithm}. Hence, the size analysis of this case will be similar to that in \cref{thm:loground_number_of_added_edges}. We now discuss the differences.

    Notice that the proof of \cref{thm:loground_number_of_added_edges} is parameterized by $p$. In our case, $p = n^{-\frac{(t+1)^{i-1}}{k}}$. Following the same analysis as in \cref{thm:loground_number_of_added_edges}, a cluster $c$ considered at iteration $j$ of epoch $i$ in expectation adds $O(1 / p)$ edges to $E_S$. Observe that $p$ is the same throughout the entire epoch. We now sum contributions of the clusters across all iterations. By applying \cref{lem:clustering_size} and the fact that a cluster proceeds from iteration $x$ to iteration $x+1$ with probability $p$, we derive
    \[
        O(1/p) \cdot \sum_{j = 1}^t O\rb{n^{1-\frac{(t+1)^{i - 1}-1}{k}}} \cdot p^{j - 1} \in O(1 / p) \cdot O\rb{n^{1-\frac{(t+1)^{i - 1}-1}{k}}} = O\rb{n^{1 + 1/k}}.
    \]

    Since there are $l$ epochs, the claim for this step follows.
\end{proof}

\remove{
\begin{proof}
A cluster $c\in \cD^{(i-1)}_t$ belongs to $\cD^{(i)}_t$ only if $c$ was sampled to $\cR^{(i)}_j$
in \cref{item:subsample_clusters} for each $1\le j\le t$.
This happens with probability $\prod\limits_{j=1}^{t} n^{-\frac{(t+1)^{i-1}}{k}} = n^{-t\cdot \frac{(t+1)^{i-1}}{k}}$.
So, the probability that a specific \emph{vertex} $v\in V^{(0)}$ survives in $V^{(i)}$ is:
\[
\prob{v\in V^{(i)}} = \prod\limits_{i'=1}^{i} n^{-t\cdot \frac{(t+1)^{(i'-1)}}{k}} = n^{-\frac{(t+1)^{i} - 1}{k}}
\implies \mathbb E\left[ |V^{(i)}|\right] = n^{1 - \frac{(t+1)^{i} - 1}{k}}
\]
\end{proof}}

Finally, we analyze the number of edges 
added in the second phase to conclude the size argument. Recall that $l = \frac{\log k}{\log(t+1)}$.
Re-phrasing \cref{cor:final_clustering_size} in the terms defined in \cref{sec:compose_clusterings},
we see that the expected number of clusters in the final $\cC^{(l)}$ is:
\[
|\cC^{(l)}| =  O\left(n^{1-\frac{(t+1)^l-1}{k}}\right) =  O\left( n^{1-\frac{k-1}{k}}\right) = O\left( n^{1/k}\right)
\]

Hence, the number of edges added in this phase is clearly upper-bound by $O(n^{1+1/k})$.
To conclude, we get the following.

\begin{theorem}
Given a weighted graph $G$ and a parameter $t$, there is an algorithm that takes  $O(\frac{t\log k}{\log(t+1)})$ iterations and outputs a spanner of size $O(n^{1+1/k} \cdot (t + \log k))$ in expectation, and stretch $O(k^s)$, where $s= \frac{\log (2t+1)}{\log(t+1)}$. 
\end{theorem}

\section{Implementation in $\MPC$ and PRAM}
\label{app:implementation}
\remove{
In this section, we state our general results for the MPC model.
\theoremmain*
For implementing our algorithms, we need to perform contractions, clustering and merging operations, and these can be implemented using  standard subroutines such as sorting, broadcasting and aggregation (\cite{goodrich2011sorting}). We explain the implementation details in \cref{app:MPC_implementation}.

\section{Omitted Proofs and Details From \cref{sec:implementation}}\label{app:MPC_implementation}
In this section show how our algorithm can be implemented in the $\MPC$ model. }

The implementation of the algorithms described is straightforward when memory per machine is $\Theta(n)$. In this case, each node along with all of its incident edges can be assigned to one machine. We can maintain a label at each edge to determine whether it is discarded or added to the spanner, and a label for each node corresponding to the largest level cluster it belongs to.
Nodes can maintain all of these information simply by communicating with their neighbors in each round, which is straightforward when memory per machine is $\Theta(n)$. Maintaining these labels and performing contractions and merges is somewhat more complicated in the low memory regime, since edges incident to a node may not fit into one machine. Hence, the low memory regime is our main focus for the rest of this section.

\subparagraph{Input Configuration.} We will create and maintain the following configuration of the input throughout the algorithm: Given a graph $G=(V,E)$, the goal is to store all the edges incident to each node $v$ in a contiguous group of machines (based on machine ID), which we denote by $M(v)$. We also denote a designated machine in the set $M(v)$ by \textit{the leader}, that represents the node. This could for instance always be the first machine in the set.  
 
 Through out the algorithm we will append each edge with certain labels (e.g., ``corresponding to a cluster'', ``corresponding to a supernode'', etc). In order to create such a setting we can use several subroutines from previous work. The following subroutines can all be implemented in $O(1/\gamma)$ rounds when memory per machine is $O(n^\gamma)$, for any constant $\gamma > 0$.
 

%
	\begin{itemize}
		\item \textsc{Sort} (\cite{goodrich2011sorting}). Given a set of $N$ comparable items, the goal is to have the items sorted on the output machines, i.e., the output machine with smaller ID holds smaller items.


		\item \textsc{Find Minimum ($v$)} (e.g., \cite{dinitz2019}). Finds the minimum value over the set of machines $M(v)$, i.e., machines containing edges incident to node $v \in V$.

		\item \textsc{Broadcast ($b, v$)} (e.g., \cite{dinitz2019}). Broadcasts a message $b$ to all machine in $M(v)$.
	\end{itemize}

Most of these subroutines can be implemented using an (implicit) aggregation tree, with branching factor $n^\gamma$, where the machines in $M(v)$ are the leaves. In each round, the aggregate function is computed over a set of $n^\gamma$ nodes in this tree and moved up to the higher levels (e.g. in case of find minimum), or moved down the tree (in case of broadcast). For more details on these standard technique, see \cite{goodrich2011sorting}.

First, using sorting and indexing subroutines we can create the configuration described above. This can be done by sorting the edges of form $(u,v)$ based on ID of the first node in the tuple. We also assume w.l.o.g that pairs $(u,v)$ and $(v,u)$ are present in the input for each undirected edge. Any time one of these tuples, say $(u,v)$ changes (or adds) a label, this label is also added to the machine containing $(v,u)$ by using the sort subroutine again but based on the \textit{smallest ID} of the two endpoints. In this case the tuples $(u,v)$ and $(v,u)$ will be place in the same machine (or machines next to each other) and the new labels can be updated for both tuples. After this process, we will sort the input again based on the ordered pairs to return to the initial configuration.

In addition to the standard subroutines described, for implementing our algorithm in $\MPC$ we also need the following subroutines:
	\begin{itemize}
		\item \textsc{Clustering:} Each node maintains a label corresponding to the highest level cluster it belongs to.

		\item \textsc{Merge:} Merging two clusters into one cluster.

		\item \textsc{Contraction:} Creating a supergraph by contracting a set of clusters to form supernodes and the corresponding superedges.
	\end{itemize}

\begin{lemma} \label{lem:subroutines}
Subroutines clustering, merge and contraction (as described above) can be implemented in $O(1/\gamma)$ rounds of $\MPC$, when memory per machine is $O(n^\gamma)$, for any constant $\gamma > 0$.
\end{lemma}
\begin{proof}
Each of these operations can be performed by creating and updating labels for each edge. In particular, we will maintain tuples of form $((u,v), \mathcal{I}_u,\mathcal{I}_v, i)$, where $i$ represents the highest level cluster center, and $\mathcal{I}_u$ represents the ID of the cluster center node $u$ belongs to. 

\textsc{Clustering}. Whenever the level of a cluster increments, we first update the level label of the cluster center. Then we can sort all the tuples by the cluster center ID of the first endpoint and increment the corresponding label. We can then repeat the process again the second endpoint. 

\textsc{Merge.} The merge operation can be performed similarly by relabeling the ID of the center for each edge that is merging into the another cluster. 


\textsc{Contraction}. For this operation, we again use the sorting subroutine to group all the nodes in a cluster to a contiguous set of machines that are simulating a supernode. We can then relabel the edge tuples so that $(u,v)$ is replaced with $(\hat{u}, \hat{v})$, where $\hat{u}$ and $\hat{v}$ are IDs of the cluster centers that were contracted. Hence each edge is replaced with a new tuple corresponding to the two supernodes that its endpoints belong to. 

At the end of each of these operations, we will sort the tuples again based on (possibly new) edge tuples to get back to the initial configuration, where all edges incident to a node (or supernode) are in adjacent. 
 All of these steps can also be done in $O(1/\gamma)$ rounds when memory per machine is $O(n^\gamma)$.
\end{proof}

We will now argue how using these subroutines, we can implement the algorithm of \cref{sec:general}, and implementation of the algorithms of \cref{sec:loground_cluster_merging} and \cref{sec:polyround_algorithm} will immediately follow.

\theoremmain*

\begin{proof}
We now describe how operation needed for a single iteration of the our algorithm can be implemented in $\MPC$ with $O(n^\gamma)$ memory per machine using the subroutines described.
 Sampling is straightforward: each cluster center will be subsampled with probability $n^{\frac{(t+1)^{i-1}}{k}}$ (at iteration $t$ of the $i$-th epoch), and using the \textsc{Clustering} subroutine, as described in \cref{lem:subroutines}, all other nodes in the subsampled cluster will update their labels.
 
 Recall that we always maintain the configuration in which the edges of each node are grouped in a contiguous set of machines at any time for the \textit{quotient graph} of each epoch. This allows us to discard edges or add edges to the spanner in a straightforward way by maintaining another label for each edge.

In \cref{item:neighboring_sampled_cluster} and \cref{item:no_neighboring_sampled_cluster}, we use the subroutine \textsc{Find Min($v$)} to find the minimum weight edge among a subset of nodes incident to a given node $v$.

After each iteration we will perform a \textsc{Merge} operation, and at the end of each epoch we will perform a \textsc{Contraction} operation. All of these operation can be performed in $O(1/\gamma)$ rounds.

Finally, for the overall number of rounds, recall that the algorithm proceeds in $\ell= \frac{\log k}{\log (t+1)}$ epoches. For each epoch we run $t$ iterations of a set of operations all of which takes $O(1/\gamma)$ rounds based on the subroutines explained.
Finally, at an additional $O(\log n)$ overhead on the overall memory, we can repeat the algorithm in parallel $O(\log n)$ rounds to turn the expected size guarantee into a high probability bound using a simple Chernoff-bounds arguments.
Hence, overall the algorithm takes $O(\frac{1}{\gamma} \cdot\frac{t\log k}{\log(t+1)})$ rounds.
\end{proof}

\subparagraph{PRAM}
Using similar primitives as \cite{baswana_sen} we can obtain low depth CRCW PRAM algorithms.
Specifically, we can use the primitives of \emph{hashing}, \emph{semisorting}, and \emph{generalized find min} that were utilized in \cite{baswana_sen}.
Additionally, we need a new primitive which has to merge two sets of vertices (clusters).
The \emph{merge} primitive can be implemented like a union find data structure,
where each set has a ``leader'' node, and all other nodes point to the leader.
The merge operation can be performed in $O(1)$ steps, by identifying and changing all the leader pointers in parallel.
The PRAM depth is the same as the MPC round complexity, with an additional multiplicative $\log^* n$ factor
that arises from the hashing, semisorting and find min primitives.
We note that $O(k)$ depth PRAM algorithms for spanners were also studied in \cite{miller2015improved}, in addition to \cite{baswana_sen}.

\section{Application for Approximate Shortest Paths} \label{sec:APSP}

While we focused on constructing spanners in a setting where each machine has a memory of sublinear size, the same algorithm also works for the less restrictive setting of $\MPC$ with linear memory, e.g., we can consider each machine as subdivided into multiple machines of less memory. The advantage in this model is that if we build a spanner of near-linear size, we can allow to store it on one machine which leads to a $poly(\log n)$-approximation for the all pairs shortest paths (APSP) problem. We next elaborate on it. 

As a reminder, our algorithm builds a spanner of size $O(n^{1+1/k} \cdot (t+\log k))$ in $poly(\log k)$ rounds. If we focus on $t=O(\log{\log{n}})$ and plug in $k=\log{n}$, we get a construction of spanner of near-linear size and $poly(\log n)$ stretch in $poly(\log{\log n})$ time. Since the spanner has size $O(n \log{\log n})$, if we allow each machine to have $\tO(n)$ memory, we can just send the whole spanner to one machine. As the spanner has paths with stretch $poly(\log n)$ between all pairs of vertices, this gives $poly(\log n)$ approximation for APSP, summarized as follows.

\appAPSP*

As two special cases, this gives $O(\log^{\log 3} n)$-approximation in $O(\log \log n)$ time, and $O(\log^{1+o(1)}n)$-approximation in $O(\log^2 \log {n})$ time.

\section{Applications in \clique} \label{sec:clique}

We next discuss applications for the closely related distributed \clique model. In this model there are $n$ nodes that communicate by sending $\Theta(\log n)$ bit messages in synchronous rounds. It was shown in \cite{DBLP:journals/corr/abs-1802-10297} that a less restricted variant of MPC, called semi-MPC is equivalent to the \clique model. This is a variant of MPC where there are $n$ machines with $\Theta(n)$ memory. Compared to the more standard MPC model, in this model the total memory is $O(n^2)$ and not $\tO(m)$. The spanner algorithm also works in this less restricted variant, with the following difference. The basic spanner algorithm gives a guarantee on the size of the spanner in expectation. To get this guarantee w.h.p, we can run it $O(\log n)$ times in parallel in the MPC model which adds logarithmic factor to the total size of memory. 
However, in the \clique model, this approach is problematic because it would multiply the complexity by an $O(\log{n})$ factor, which is too expensive as we aim for a $poly(\log{\log{n}})$ complexity.
To avoid this, we next sketch how to do parallel repetitions in the \clique efficiently, which results in a similar spanner construction for the \clique model.

\begin{theorem}
There is an algorithm that runs in $O(\frac{t\log k}{\log(t+1)})$ rounds of \clique and w.h.p.~outputs a spanner of size $O(n^{1+1/k} \cdot (t + \log k))$, and stretch $O(k^s)$, where $s= \frac{\log (2t+1)}{\log(t+1)}$. 
\end{theorem}

\begin{proof}
As discussed above, we can run one run of the spanner algorithm in the semi-MPC model, and hence also in the \clique model using the simulation described in \cite{DBLP:journals/corr/abs-1802-10297}.
This gives $O(k^s)$-spanner of expected size $O(n^{1+1/k} \cdot (t + \log k))$ in $O(\frac{t\log k}{\log(t+1)})$ rounds. 
We next explain how to run $O(\log n)$ runs of the algorithm in parallel, which would enable us to build a spanner of size $O(n^{1+1/k} \cdot (t + \log k))$ w.h.p. 

The size analysis of the spanner is based on the following randomized process. In each iteration\footnote{This also refers to iterations where we apply Baswana-Sen, as it is based on the same randomized process.} of the algorithm, we have a set of super-nodes $C$, and we sample from them a set of clusters $R$ such that each vertex in $C$ is added to $R$ with probability $p$. The size analysis is based on the following: 
\begin{enumerate}
\item The expected number of clusters is $O(|C|p)$.\label{clusters_number}
\item The expected number of edges added to the spanner in this iteration is $O(|C|/p)$.\label{spanner_edges}
\end{enumerate}
A standard application of Chernoff bound shows that \ref{clusters_number} actually holds w.h.p (as long as the number of clusters is $\Omega(\log{n})$, but when it is smaller it is easy to show that the total number of edges added to the spanner is small enough). Also, from Markov's inequality, it follows that \ref{spanner_edges} happens with constant probability. It follows that if we simulate this randomized process $O(\log{n})$ times in parallel, we get w.h.p a run where both events hold, which implies that we get the bound on the number of edges w.h.p. Hence, we just need to simulate this part $O(\log{n})$ times (in each iteration), to guarantee a total of $O(n^{1+1/k} \cdot (t + \log k))$ edges w.h.p. 

This is done as follows. In the algorithm, first, each super-node from $C$ is added to $R$ with probability $p$. The sampling is a completely local process done by each super-node separately. Hence, each super-node can just simulate it $O(\log{n})$ times. Then, we would like each super-node to let all other nodes learn the outcome of the sampling process. This can be done by just sending a single $O(\log{n})$ bit message from each super-node $v$ to all nodes, where the $i$'th bit of the message indicates if $v$ is added to $R$ in the $i$'th run of the algorithm. This takes only one round. Now, based on this information, each super-node in $C$ knows exactly which edges adjacent to it are taken to the spanner in each run of the algorithm. To conclude, we choose $O(\log{n})$ nodes that would be responsible for the $O(\log{n})$ different runs. Node $v_i$ would collect from all super-nodes the number of edges adjacent to them added to the spanner in run $i$. In addition, all nodes know how many vertices were added to $R$ in each one of the runs. Finally, we choose a run where the number of nodes in $R$ is $O(|C|p)$, and the number of edges added to the spanner is $O(|C|/p)$, which exists w.h.p as explained above. We only added a constant number of rounds for each iteration, hence the time complexity matches the complexity of the MPC algorithm.
\end{proof}

\subparagraph{Approximate Shortest Paths.}
The spanner algorithm results in a $poly(\log{n})$-approximation for weighted APSP in $poly(\log{\log{n}})$ rounds in the \clique model, as follows. 
Choosing $k=\log{n}$ and $t=O(\log{\log{n}})$, results in a spanner of size $O(n\log{\log{n}})$. We can let all vertices learn the whole spanner in $O(\log{\log{n}})$ rounds using Lenzen's routing \cite{lenzen2013optimal}, which results in the following corollary.

\APSPclique*
\section*{Acknowledgements}
We thank Merav Parter, Michael Dinitz and Aditya Krishnan for fruitful discussions.

\paragraph{Funding.} AS.~Biswas is supported by MIT-IBM Watson AI Lab and research collaboration agreement No. W1771646, NSF awards CCF-1733808, IIS-1741137, Big George Ventures Fund Fellowship. 
M.~Dory is supported in part by the Swiss National
Foundation No. $200021\_184735$.
M.~Ghaffari is supported in part by the European Research Council (ERC) under the European Union's Horizon 2020 research and innovation program No. 853109.
S.~Mitrovi\'c is supported by the Swiss NSF grant No. $P400P2\_191122/1$,  MIT-IBM Watson AI Lab and research collaboration agreement No. W1771646, NSF award CCF-1733808, and FinTech@CSAIL. 
Y.~Nazari is supported by NSF award CCF-190911.

\bibliographystyle{alpha}
\bibliography{ref}

\newcommand{\etalchar}[1]{$^{#1}$}
\begin{thebibliography}{CHDKL19}

\bibitem[ABB{\etalchar{+}}19]{AssadiBBMS19}
Sepehr Assadi, MohammadHossein Bateni, Aaron Bernstein, Vahab~S. Mirrokni, and
  Cliff Stein.
\newblock Coresets meet {EDCS:} algorithms for matching and vertex cover on
  massive graphs.
\newblock In Timothy~M. Chan, editor, {\em Proc.~Symposium on Discrete
  Algorithms (SODA)}, pages 1616--1635, 2019.

\bibitem[ACK19]{AssadiCK18}
Sepehr {Assadi}, Yu~{Chen}, and Sanjeev {Khanna}.
\newblock {Sublinear Algorithms for $(\Delta + 1)$ Vertex Coloring}.
\newblock In {\em Proceedings 30th Annual {ACM-SIAM} Symposium on Discrete
  Algorithms ({SODA})}, 2019.

\bibitem[AG15]{AhnGuha15}
Kook~Jin Ahn and Sudipto Guha.
\newblock Access to data and number of iterations: Dual primal algorithms for
  maximum matching under resource constraints.
\newblock In {\em SPAA}, pages 202--211, 2015.

\bibitem[AGM12]{AhnGM12}
K.~J. Ahn, S.~Guha, and A.~McGregor.
\newblock Analyzing graph structure via linear measurements.
\newblock In {\em Proceedings of the 23rd Annual ACM-SIAM Symposium on Discrete
  Algorithms (SODA)}, pages 459--467, 2012.

\bibitem[ANOY14]{Andoni:2014}
Alexandr Andoni, Aleksandar Nikolov, Krzysztof Onak, and Grigory Yaroslavtsev.
\newblock Parallel algorithms for geometric graph problems.
\newblock In {\em Proc.~Symposium on Theory of Computation (STOC)}, pages
  574--583, 2014.

\bibitem[Ass17]{assadi2017simple}
Sepehr Assadi.
\newblock Simple round compression for parallel vertex cover.
\newblock {\em arXiv preprint arXiv:1709.04599}, 2017.

\bibitem[ASS{\etalchar{+}}18]{Andoni2018}
Alexandr Andoni, Clifford Stein, Zhao Song, Zhengyu Wang, and Peilin Zhong.
\newblock Parallel graph connectivity in log diameter rounds.
\newblock In {\em Proc.~Foundations of Computer Science (FOCS)}, pages
  674--685, 2018.

\bibitem[ASW19]{AssadiSW19}
Sepehr Assadi, Xiaorui Sun, and Omri Weinstein.
\newblock Massively parallel algorithms for finding well-connected components
  in sparse graphs.
\newblock In Peter Robinson and Faith Ellen, editors, {\em Proc.~Principles of
  Distributed Computing (PODC)}, pages 461--470, 2019.

\bibitem[ASZ19]{AndoniSZ19}
Alexandr Andoni, Clifford Stein, and Peilin Zhong.
\newblock Log diameter rounds algorithms for 2-vertex and 2-edge connectivity.
\newblock In Christel Baier, Ioannis Chatzigiannakis, Paola Flocchini, and
  Stefano Leonardi, editors, {\em Proc.~ICALP}, volume 132, pages 14:1--14:16,
  2019.

\bibitem[ASZ20]{AndoniCZ20}
Alexandr Andoni, Clifford Stein, and Peilin Zhong.
\newblock Parallel approximate undirected shortest paths via low hop emulators.
\newblock In {\em Proc.~Symposium on Theory of Computation (STOC)}, 2020.

\bibitem[BBD{\etalchar{+}}19]{Behnezhad0DFHKU19}
Soheil Behnezhad, Sebastian Brandt, Mahsa Derakhshan, Manuela Fischer,
  MohammadTaghi Hajiaghayi, Richard~M. Karp, and Jara Uitto.
\newblock Massively parallel computation of matching and {MIS} in sparse
  graphs.
\newblock In Peter Robinson and Faith Ellen, editors, {\em Proc.~Principles of
  Distributed Computing (PODC)}, pages 481--490, 2019.

\bibitem[BDE{\etalchar{+}}19]{BehnezhadDELM19}
Soheil Behnezhad, Laxman Dhulipala, Hossein Esfandiari, Jakub Lacki, and
  Vahab~S. Mirrokni.
\newblock Near-optimal massively parallel graph connectivity.
\newblock In David Zuckerman, editor, {\em Proc.~Foundations of Computer
  Science (FOCS)}, pages 1615--1636, 2019.

\bibitem[BDH18]{DBLP:journals/corr/abs-1802-10297}
Soheil Behnezhad, Mahsa Derakhshan, and MohammadTaghi Hajiaghayi.
\newblock Brief announcement: Semi-mapreduce meets congested clique.
\newblock {\em CoRR}, abs/1802.10297, 2018.

\bibitem[BEG{\etalchar{+}}18]{boroujeni2018approximating}
Mahdi Boroujeni, Soheil Ehsani, Mohammad Ghodsi, MohammadTaghi HajiAghayi, and
  Saeed Seddighin.
\newblock Approximating edit distance in truly subquadratic time: quantum and
  {M}ap{R}educe.
\newblock In {\em Proc.~Symposium on Discrete Algorithms (SODA)}, pages
  1170--1189, 2018.

\bibitem[BFU18]{brandt2018breaking}
Sebastian Brandt, Manuela Fischer, and Jara Uitto.
\newblock Breaking the linear-memory barrier in {MPC}: Fast {MIS} on trees with
  $n^{\epsilon}$ memory per machine.
\newblock {\em arXiv preprint arXiv:1802.06748}, 2018.

\bibitem[BHH19]{BehnezhadHH19}
Soheil Behnezhad, MohammadTaghi Hajiaghayi, and David~G. Harris.
\newblock Exponentially faster massively parallel maximal matching.
\newblock In David Zuckerman, editor, {\em Proc.~Foundations of Computer
  Science (FOCS)}, pages 1637--1649, 2019.

\bibitem[BKKL17]{becker_et_al:transshipment}
Ruben Becker, Andreas Karrenbauer, Sebastian Krinninger, and Christoph Lenzen.
\newblock {Near-Optimal Approximate Shortest Paths and Transshipment in
  Distributed and Streaming Models}.
\newblock In {\em Proc.~Symposium on DIStributed Computing (DISC)}, volume~91,
  pages 7:1--7:16, 2017.

\bibitem[BKS13]{Beame13}
Paul Beame, Paraschos Koutris, and Dan Suciu.
\newblock Communication steps for parallel query processing.
\newblock In {\em Proceedings of the 32Nd ACM SIGMOD-SIGACT-SIGAI Symposium on
  Principles of Database Systems (PODS)}, pages 273--284, 2013.

\bibitem[BKS14]{Beame14}
Paul Beame, Paraschos Koutris, and Dan Suciu.
\newblock Skew in parallel query processing.
\newblock In {\em Proceedings of the 33rd ACM SIGMOD-SIGACT-SIGART Symposium on
  Principles of Database Systems (PODS)}, pages 212--223, 2014.

\bibitem[BLP20]{ben2020new}
Uri Ben-Levy and Merav Parter.
\newblock New ($\alpha$, $\beta$) spanners and hopsets.
\newblock In {\em Proceedings of the Fourteenth Annual ACM-SIAM Symposium on
  Discrete Algorithms}, pages 1695--1714. SIAM, 2020.

\bibitem[BS07]{baswana_sen}
Surender Baswana and Sandeep Sen.
\newblock A simple and linear time randomized algorithm for computing sparse
  spanners in weighted graphs.
\newblock {\em Random Structures \& Algorithms}, 30(4):532--563, 2007.

\bibitem[CFG{\etalchar{+}}19]{ChangFGUZ19}
Yi{-}Jun Chang, Manuela Fischer, Mohsen Ghaffari, Jara Uitto, and Yufan Zheng.
\newblock The complexity of ({\(\Delta\)}+1) coloring in congested clique,
  massively parallel computation, and centralized local computation.
\newblock In {\em Proc.~Principles of Distributed Computing (PODC)}, pages
  471--480, 2019.

\bibitem[CHDKL19]{censor2019fast}
Keren Censor-Hillel, Michal Dory, Janne~H Korhonen, and Dean Leitersdorf.
\newblock Fast approximate shortest paths in the congested clique.
\newblock In {\em Proceedings of the 2019 ACM Symposium on Principles of
  Distributed Computing (PODC)}, pages 74--83, 2019.

\bibitem[CLM{\etalchar{+}}18]{czumaj2017round}
Artur Czumaj, Jakub Lacki, Aleksander Madry, Slobodan Mitrovic, Krzysztof Onak,
  and Piotr Sankowski.
\newblock Round compression for parallel matching algorithms.
\newblock In {\em Proc.~Symposium on Theory of Computation (STOC)}, pages
  471--484, 2018.

\bibitem[Coh00]{cohen2000polylog}
Edith Cohen.
\newblock Polylog-time and near-linear work approximation scheme for undirected
  shortest paths.
\newblock {\em Journal of the ACM (JACM)}, 47(1):132--166, 2000.

\bibitem[DG08]{dean2008mapreduce}
Jeffrey Dean and Sanjay Ghemawat.
\newblock Mapreduce: simplified data processing on large clusters.
\newblock {\em Communications of the ACM}, 51(1):107--113, 2008.

\bibitem[DGPV08]{derbel2008locality}
Bilel Derbel, Cyril Gavoille, David Peleg, and Laurent Viennot.
\newblock On the locality of distributed sparse spanner construction.
\newblock In {\em Proc.~Principles of Distributed Computing (PODC)}, pages
  273--282, 2008.

\bibitem[DN19]{dinitz2019}
Michael Dinitz and Yasamin Nazari.
\newblock Massively parallel approximate distance sketches.
\newblock In {\em Proceedings of International Conference on Principles of
  Distributed Systems (OPODIS 2019)}. Schloss Dagstuhl-Leibniz-Zentrum fuer
  Informatik, 2019.

\bibitem[DP20]{CliqueShortestPaths2020}
Michal Dory and Merav Parter.
\newblock Exponentially faster shortest paths in the congested clique.
\newblock In {\em PODC}, 2020.

\bibitem[FKN21]{filtser2021}
Arnold Filtser, Michael Kapralov, and Navid Nouri.
\newblock Graph spanners by sketching in dynamic streams and the simultaneous
  communication model.
\newblock In {\em Proceedings of the 2021 ACM-SIAM Symposium on Discrete
  Algorithms (SODA)}. SIAM, 2021.

\bibitem[FL18]{friedrichs2018}
Stephan Friedrichs and Christoph Lenzen.
\newblock Parallel metric tree embedding based on an algebraic view on
  moore-bellman-ford.
\newblock {\em Journal of the ACM (JACM)}, 65(6):43, 2018.

\bibitem[GGK{\etalchar{+}}18]{ghaffari2018improved}
Mohsen Ghaffari, Themis Gouleakis, Christian Konrad, Slobodan Mitrovi{\'c}, and
  Ronitt Rubinfeld.
\newblock Improved massively parallel computation algorithms for mis, matching,
  and vertex cover.
\newblock In {\em Proc.~Principles of Distributed Computing (PODC)}.
  arXiv:1802.08237, 2018.

\bibitem[GKMS19]{GamlathKMS19}
Buddhima Gamlath, Sagar Kale, Slobodan Mitrovic, and Ola Svensson.
\newblock Weighted matchings via unweighted augmentations.
\newblock In Peter Robinson and Faith Ellen, editors, {\em Proc.~Principles of
  Distributed Computing (PODC)}, pages 491--500. {ACM}, 2019.

\bibitem[GKU19]{GhaffariKU19}
Mohsen Ghaffari, Fabian Kuhn, and Jara Uitto.
\newblock Conditional hardness results for massively parallel computation from
  distributed lower bounds.
\newblock In {\em Proc.~Foundations of Computer Science (FOCS)}, pages
  1650--1663. {IEEE} Computer Society, 2019.

\bibitem[GLM19]{GhaffariLM19}
Mohsen Ghaffari, Silvio Lattanzi, and Slobodan Mitrovic.
\newblock Improved parallel algorithms for density-based network clustering.
\newblock In Kamalika Chaudhuri and Ruslan Salakhutdinov, editors, {\em
  Proceedings of the International Conference on Machine Learning (ICML)},
  volume~97, pages 2201--2210. {PMLR}, 2019.

\bibitem[GN20]{GhaffariN20-cut}
Mohsen Ghaffari and Krzysztof Nowicki.
\newblock {Massively Parallel Algorithms for Minimum Cut}.
\newblock In {\em Proc.~Principles of Distributed Computing (PODC)}, page to
  appear, 2020.

\bibitem[GNT20]{GhaffariNT20-cut}
Mohsen Ghaffari, Krzysztof Nowicki, and Mikkel Thorup.
\newblock {Faster Algorithms for Edge Connectivity via Random 2-Out
  Contractions}.
\newblock In {\em Proc.~Symposium on Discrete Algorithms (SODA)}, pages
  1260--1279, 2020.

\bibitem[GSZ11]{goodrich2011sorting}
Michael~T. Goodrich, Nodari Sitchinava, and Qin Zhang.
\newblock Sorting, searching, and simulation in the {MapReduce} framework.
\newblock In {\em Proc.~ISAAC}, pages 374--383. Springer, 2011.

\bibitem[GU19]{GhaffariU19}
Mohsen Ghaffari and Jara Uitto.
\newblock Sparsifying distributed algorithms with ramifications in massively
  parallel computation and centralized local computation.
\newblock In {\em Proc.~Symposium on Discrete Algorithms (SODA)}, pages
  1636--1653, 2019.

\bibitem[HLL18]{harvey2018greedy}
Nicholas J.~A. Harvey, Christopher Liaw, and Paul Liu.
\newblock Greedy and local ratio algorithms in the {MapReduce} model.
\newblock In {\em Proceedings of the 30th on Symposium on Parallelism in
  Algorithms and Architectures {(SPAA)}}, pages 43--52, New York, NY, USA,
  2018. ACM.

\bibitem[HLSS19]{hajiaghayi2019mapreduce}
MohammadTaghi Hajiaghayi, Silvio Lattanzi, Saeed Seddighin, and Cliff Stein.
\newblock Mapreduce meets fine-grained complexity: Mapreduce algorithms for
  apsp, matrix multiplication, 3-sum, and beyond.
\newblock {\em arXiv preprint arXiv:1905.01748}, 2019.

\bibitem[HP15]{hegeman2015lessons}
James~W Hegeman and Sriram~V Pemmaraju.
\newblock Lessons from the congested clique applied to {M}ap{R}educe.
\newblock {\em Theoretical Computer Science}, 608:268--281, 2015.

\bibitem[IBY{\etalchar{+}}07]{isard2007dryad}
Michael Isard, Mihai Budiu, Yuan Yu, Andrew Birrell, and Dennis Fetterly.
\newblock Dryad: distributed data-parallel programs from sequential building
  blocks.
\newblock In {\em Proceedings of the 2nd ACM SIGOPS/EuroSys European Conference
  on Computer Systems 2007}, pages 59--72, 2007.

\bibitem[ILMP19]{ItalianoLMP19}
Giuseppe~F. Italiano, Silvio Lattanzi, Vahab~S. Mirrokni, and Nikos Parotsidis.
\newblock Dynamic algorithms for the massively parallel computation model.
\newblock In {\em SPAA}, pages 49--58, 2019.

\bibitem[IMS17]{Im17}
Sungjin Im, Benjamin Moseley, and Xiaorui Sun.
\newblock Efficient massively parallel methods for dynamic programming.
\newblock In {\em Proc.~Symposium on Theory of Computation (STOC)}, pages
  798--811, 2017.

\bibitem[KSV10]{karloff2010MapReduce}
Howard Karloff, Siddharth Suri, and Sergei Vassilvitskii.
\newblock A model of computation for {M}ap{R}educe.
\newblock In {\em Proc.~Symposium on Discrete Algorithms (SODA)}, pages
  938--948, 2010.

\bibitem[KW14]{kapralov2014}
Michael Kapralov and David Woodruff.
\newblock Spanners and sparsifiers in dynamic streams.
\newblock In {\em Proceedings of the 2014 ACM symposium on Principles of
  distributed computing}, pages 272--281, 2014.

\bibitem[Len13]{lenzen2013optimal}
Christoph Lenzen.
\newblock Optimal deterministic routing and sorting on the congested clique.
\newblock In {\em Proceedings of the 2013 ACM symposium on Principles of
  distributed computing (PODC)}, pages 42--50, 2013.

\bibitem[Li20]{li2020_PRAM}
Jason Li.
\newblock Faster parallel algorithm for approximate shortest path.
\newblock In {\em Proc.~Symposium on Theory of Computation (STOC)}, 2020.

\bibitem[LMOS20]{LackiMOS20}
Jakub Lacki, Slobodan Mitrovic, Krzysztof Onak, and Piotr Sankowski.
\newblock Walking randomly, massively, and efficiently.
\newblock In {\em Proc.~Symposium on Theory of Computation (STOC)}, 2020.

\bibitem[LMSV11]{LattanziMSV11}
Silvio Lattanzi, Benjamin Moseley, Siddharth Suri, and Sergei Vassilvitskii.
\newblock Filtering: a method for solving graph problems in {MapReduce}.
\newblock In {\em SPAA}, pages 85--94, 2011.

\bibitem[MPVX15]{miller2015improved}
Gary~L Miller, Richard Peng, Adrian Vladu, and Shen~Chen Xu.
\newblock Improved parallel algorithms for spanners and hopsets.
\newblock In {\em Proceedings of the 27th ACM symposium on Parallelism in
  Algorithms and Architectures}, pages 192--201, 2015.

\bibitem[NMN01]{Boruvka}
Jaroslav Ne{\v{s}}et{\v{r}}il, Eva Milkov{\'a}, and Helena
  Ne{\v{s}}et{\v{r}}ilov{\'a}.
\newblock Otakar boruvka on minimum spanning tree problem translation of both
  the 1926 papers, comments, history.
\newblock {\em Discrete Mathematics}, 233(1):3--36, 2001.

\bibitem[Pel00]{Peleg00}
David Peleg.
\newblock {\em Distributed Computing: A Locality-Sensitive Approach}.
\newblock SIAM, 2000.

\bibitem[PS89]{peleg1989graph}
David Peleg and Alejandro~A Sch{\"a}ffer.
\newblock Graph spanners.
\newblock {\em Journal of graph theory}, 13(1):99--116, 1989.

\bibitem[PY18]{parter2018spanner}
Merav Parter and Eylon Yogev.
\newblock Congested clique algorithms for graph spanners.
\newblock In {\em 32nd International Symposium on Distributed Computing (DISC
  2018)}. Schloss Dagstuhl-Leibniz-Zentrum fuer Informatik, 2018.

\bibitem[RG20]{RozhonGhaffari20}
Vaclav Rozhon and Mohsen Ghaffari.
\newblock Polylogarithmic-time deterministic network decomposition and
  distributed derandomization.
\newblock In {\em Proc.~Symposium on Theory of Computation (STOC)}, 2020.

\bibitem[RVW16]{Roughgarden16}
Tim Roughgarden, Sergei Vassilvitskii, and Joshua~R. Wang.
\newblock Shuffles and circuits: (on lower bounds for modern parallel
  computation).
\newblock In {\em SPAA}, pages 1--12, 2016.

\bibitem[Whi12]{white2012hadoop}
Tom White.
\newblock {\em Hadoop: The definitive guide}.
\newblock " O'Reilly Media, Inc.", 2012.

\bibitem[ZCF{\etalchar{+}}10]{ZahariaCFSS10}
Matei Zaharia, Mosharaf Chowdhury, Michael~J. Franklin, Scott Shenker, and Ion
  Stoica.
\newblock Spark: Cluster computing with working sets.
\newblock In {\em 2nd {USENIX} Workshop on Hot Topics in Cloud Computing
  (HotCloud)}, 2010.

\end{thebibliography}

\appendix
\section{Related Work on MPC algorithms for Other Graph Problems}
\label{app:related}
Here, we provide a brief and high-level overview of the recent progress on massively parallel algorithms for graph problems. We discuss these in two categories:

One category is the problem of connected components and other related problems such as maximal forest, minimum-weight spanning tree, and minimum cut (in unweighted graphs). For the super-linear and even near-linear memory regimes, there are $O(1)$ round algorithms~\cite{LattanziMSV11, AhnGM12} for connected components, maximal forest, and minimum weight spanning tree. For the minimum cut problem, there are now $O(1)$ round algorithms for both the unweighted~\cite{GhaffariNT20-cut} and also weighted~\cite{GhaffariN20-cut} graphs, using near-linear memory. 
In contrast, for the strongly sublinear memory regime, even for the connected components problem, the best known is an $O(\log n)$ time algorithm~\cite{karloff2010MapReduce} that follows from Boruvka's classic method~\cite{Boruvka}, and this complexity is widely believed to be optimal.\footnote{In fact often used as a hardness assumption for other conditional lower bounds, see e.g., \cite{GhaffariKU19,LackiMOS20}. It is known that this $\Omega(\log n)$ complexity lower bounds for many natural algorithmic approaches~\cite{Beame13} but that proving it as a general $\Omega(\log n)$ lower bound would lead to a significant breakthrough in circuit complexity, refuting that $P\subset NC_1$~\cite{Roughgarden16}} Recent work has improved the complexity for connected components to $O(\log D + \log\log n)$, where $D$ denotes the graph diameter (per connected component)~\cite{Andoni2018,BehnezhadDELM19}. 

Another category is problems with a local nature (in the style of those studied frequently in the LOCAL model as symmetry breaking tasks), including maximum matching approximation, graph coloring with $\Delta+1$ colors where $\Delta$ denotes the maximum degree, maximal independent set, vertex cover approximation, etc. For all of these, there are $O(1)$ algorithms in the super-linear memory regime~\cite{LattanziMSV11, harvey2018greedy}. A recent line of work brought the complexity down to $O(\log\log n)$ also in the near-linear memory regime, for maximal matching approximation, vertex cover approximation and maximal independent set~\cite{czumaj2017round,ghaffari2018improved,AssadiBBMS19}, and to $O(1)$ for coloring~\cite{AssadiCK18,ChangFGUZ19}. There has been also some progress in the strongly sublinear memory regime, with $\tilde{O}(\sqrt{\log n})$ complexity for all of these problems except coloring~\cite{GhaffariU19}, and $\Omega(\log\log n)$ round conditional lower bounds are known for them, conditioned on the aforementioned $\Omega(\log n)$ complexity of the connectivity problem~\cite{GhaffariKU19}. Finally, for the $\Delta+1$ coloring problem, the state of the art is much faster and stands at $O(\log\log\log n)$ rounds~\cite{ChangFGUZ19,RozhonGhaffari20}.

\section{Spanners with $O(k)$-stretch in Suboptimal Total Memory}
\label{sec:O(k)-spanners}


In this section we discuss how one can extend the ideas\footnote{Mainly from Section~5.3 in \url{https://drops.dagstuhl.de/opus/volltexte/2018/9829/pdf/LIPIcs-DISC-2018-40.pdf}.} of \cite{parter2018spanner} to prove \cref{thm:super-linear-memory}.

\subsection{A Brief Overview of \cite{parter2018spanner}}
On a high-level, the algorithm of \cite{parter2018spanner} splits the vertices into \emph{sparse} and \emph{dense}. A vertex $v$ is called sparse if its $(k/2)$-hop neighborhood has size $O(\sqrt{n})$, and otherwise it is called dense.

The idea now is to for each sparse vertex collect its $(k/2)$-hop neighborhood and simulate there any distributed algorithm, e.g., \cite{baswana_sen}. This effectively enables us to handle sparse vertices. To handle dense vertices, we contract them in such a way that after the contraction one obtains a graph on $n^{\alpha}$ vertices, for some constant $\alpha < 1$. 

The entire process is now repeated on the contracted graph recursively for $O(1)$ times, until the graph becomes small enough. The authors show that in this way is obtained a spanner of size $O(k n^{1 + 1/k})$ and $O(k)$ stretch. Moreover, this approach requires $O(\log{k})$ rounds in Congested Clique.

\subsection{Our Adaptation to $\MPC$}
Collecting balls of size $O(\sqrt{n})$ around each vertex would require a total memory of $O(m + n^{3/2})$. To reduce it to $O(m + n^{1 + \gamma})$, we collect balls of size $O(n^{\gamma / 2})$ (including both edges and vertices) and define sparse and dense accordingly. We will ensure that for a dense vertex its ball has size $\Theta(n^{\gamma / 2})$. This property is important as it directly implies that the collected ball of a dense vertex contains $\Omega(n^{\gamma / 4})$ vertices.

\subparagraph{Sparse vertices.} Since our goal is to obtain a $O(k)$-stretch spanner, we can afford to collect larger than $(k/2)$-hop neighborhoods and simulate an algorithm locally there. In our modification, we collect $(4k)$-hop neighborhoods (or only $\Theta(n^{\gamma/2})$-size connected part of it if the size exceeds $O(n^{\gamma/2})$).\footnote{A similar idea was observed in \cite{parter2018spanner}, see Footnote~5 therein.} Then, for each sparse vertex $v$ we simulate $k$ iterations of \cite{baswana_sen} for $v$ and for each vertex in its $(k + 1)$-hop neighborhood and add the corresponding edges to the spanner. Let $e$ be an edge incident to $v$. In this way, all the vertices which are at distance at most $k$ from $e$ will locally simulate \cite{baswana_sen} and hence correctly decide whether $e$ should be added to the spanner or not. If $e$ is not added to the spanner, then this approach will properly span $e$. Overall, this allows us to properly add to the spanner each edge incident to each sparse vertex. Notice that this simulation is local and requires no additional rounds. In \cref{sec:super-linear-memory-implementation} we discuss how to collect $O(k)$-hop neighborhoods in $\MPC$.

To have consistent simulation of \cite{baswana_sen} over distinct balls, we use the same randomness for each vertex. Notice that each vertex requires only $O(k \log{n})$ random bits for simulation, which can be assigned to each vertex at the beginning of the algorithm. As long as $k \log{n} \in O(n^{\gamma/2})$, replicating the necessary randomness over all $n$ balls in total requires $O(n^{1 + \gamma})$ memory.

\subparagraph{Dense vertices.} So, it remains to properly handle edges that connect dense vertices. We also use a hitting set similar to how it is used in \cite{parter2018spanner}. Namely, we choose $\tO(n^{1 - \gamma / 4})$ vertices uniformly at random. Let that set be $Z$. This ensures that with high probability each dense ball is hit by at least one vertex from $Z$; recall that a dense ball contains $\Omega(n^{\gamma/4})$ vertices. Now, for each heavy vertex $v$ we add to the spanner a shortest path between $v$ and one of the vertices $w \in Z$ in its ball. This adds $O(k n)$ edges. We say that $v$ is \emph{assigned} to $w$. Observe that in this way any edge connecting two heavy vertices that are assigned to the same vertex from $Z$ has stretch $O(k)$. So, it remains to handle edges that have heavy endpoints such that the endpoints are assigned to distinct vertices of $Z$.

We consider a graph only on the vertices of $Z$. Two vertices $u, v \in Z$ are connected only if there exists a pair of vertices $x$ and $y$ such that $x$ and $y$ are adjacent in $G$, and $x$ is assigned to $u$ while $y$ is assigned to $v$. This graph has $\tO\rb{n^{1 - \gamma/4}}$ vertices. We now compute a $(4 / \gamma)$-stretch spanner on that graph (by simulating iteration by iteration of \cite{baswana_sen}), which has size $\tO\rb{n^{(1 - \gamma/4) (1 + \gamma / 4)}} \in O(n)$. We properly map back to $G$ the edges in this spanner, which ensures $O(k / \gamma)$ stretch of edges with endpoints in heavy vertices which are assigned to distinct vertices of $Z$. Recall that we assume that $\gamma$ is a constant.

This concludes the algorithm description.

\subsubsection{$\MPC$ Implementation}
\label{sec:super-linear-memory-implementation}
Next, we explain how the exponentiation is performed, while taking into account sparse and dense balls. A standard graph exponentiation was used in the context of $\MPC$ already, e.g., see~\cite{GhaffariLM19,GhaffariU19}. On a high-level, in step $i$ each vertex holds a ball of radius $2^i$, and in the next step attempts to obtain a ball of radius $2^{i + 1}$. Let $B_i(v)$ be the set of vertices that vertex $v$ at the beginning of iteration $i$ has. To obtain $B_{i + 1}(v)$, $v$ ``requests'' each vertex $w \in B_i(v)$ for $B_i(w)$.

Notice that here the number of requests for any ball $B_i(\cdot)$ is upper-bounded by the size of the largest ball. So, if we can fit each ball in the local memory, it means that the number of requests will be relatively small, e.g., $O(n^{\gamma})$ or even $O(n^{\gamma/ 2})$ with a proper setting of parameters. However, this property does not hold when we have both dense and sparse vertices, and where our goal is to stop growing dense balls. For instance, consider the star graph with the center in $c$. Vertex $c$ becomes dense immediately (and $B_0(c)$ will contain $O(n^{\gamma/2})$ of some of the neighbors of $c$), while the leaves are sparse in the first iteration. However, already in the first iteration $c$ receives $n - 1$ requests, which is significantly larger than $n^{\gamma/2}$.

Nevertheless, the total number of requests in the whole graph is $O(n^{1 + \gamma/2})$ and each request needs to fetch a ball of size $O(n^{\gamma / 2})$. This in total requires $O(n^{1 + \gamma})$ messages, which fits into our framework. Now, to satisfy all requests for a given vertex $w$, we first sort all the requests, $\Theta(n^{\gamma / 2})$ requests per machine. All the requests asking for $B(w)$ will appear on consecutive machines. Now, we can propagate $B(w)$ to all those machine by building a $\Theta(n^{\gamma / 2})$-ary tree on top of those machines. In this way, each machine receives $O(n^{\gamma/2} \cdot |B(w)|) \in O(n^{\gamma})$ messages. Once those machines receive $B(w)$, they each send $B(w)$ to machines/vertices that requested $B(w)$.

As a reminder, our goal is also to ensure that balls of dense vertices have size $\Theta(n^{\gamma/2})$, i.e., to ensure that such balls are not too small. To achieve that, consider a vertex $v$ that receives balls for all the requests it created and it happens that they result in a dense ball. Then, $v$ will start performing BFS within that dense aggregated ball, and terminate the exploration as soon as the size exceeds $a \cdot n^{\gamma / 2}$ for some fixed constant $a$.

Other steps of the described algorithm are straightforward to implement in $\MPC$ by using some of the standard techniques. This concludes the $\MPC$ implementation.




\end{document}